\documentclass[preprint,authoryear,11pt]{elsarticle}

\usepackage{natbib}
\usepackage{epsfig}
\usepackage{graphicx}
\usepackage{bm}
\usepackage{amsmath,amssymb,amsfonts,mathrsfs,textcomp,mathtools}

\let\notORI\not 
\usepackage[mathb]{mathabx}
\let\not\notORI

\usepackage{url}
\usepackage{multirow}
\usepackage{float}
\usepackage{aas_macros}

\usepackage{siunitx}
\graphicspath{{figures/}}

\newcommand{\SNR}{\text{SNR}}

\newcommand{\yarko}[1]{87}       
\newcommand{\rejected}[1]{14}    
\newcommand{\marginal}[1]{24}    
\newcommand{\snrh}[1]{101}       
\newcommand{\snrl}[1]{437}       
\newcommand{\srp}[1]{four}       
\newcommand{\srptot}[1]{10}      
\newcommand{\nosignal}[1]{85\%}  

\newcommand{\yarkojpl}[1]{89}

\newcommand{\JPL}{\textsc{jpl}}

\journal{Astronomy \& Astrophysics}

\begin{document}

\title{%
  Detecting the Yarkovsky effect among near-Earth asteroids from
  astrometric data%
}

\author[pi,sds]{Alessio Del Vigna}
\ead{delvigna@mail.dm.unipi.it}
\author[neocc]{Laura Faggioli}
\author[pi]{Andrea Milani}
\author[imcce]{Federica Spoto}
\author[jpl]{Davide Farnocchia}
\author[oca]{Benoit Carry}

\address[pi]{Dipartimento di Matematica, Universit\`a di Pisa, Largo
  Bruno Pontecorvo 5, Pisa, Italy}
\address[sds]{Space Dynamics Services s.r.l., via Mario Giuntini,
  Navacchio di Cascina, Pisa, Italy}
\address[imcce]{IMCCE, Observatoire de Paris, PSL Research University,
  CNRS, Sorbonne Universités, UPMC Univ. Paris 06, Univ. Lille, 77
  av. Denfert-Rochereau F-75014 Paris, France}
\address[neocc]{ESA SSA-NEO Coordination Centre, Largo Galileo
  Galilei, 1, 00044 Frascati (RM), Italy}
\address[jpl]{Jet Propulsion Laboratory/California Institute of
  Technology, 4800 Oak Grove Drive, Pasadena, 91109 CA, USA}
\address[oca]{Universit\'e C\^ote d'Azur, Observatoire de la C\^ote
  d'Azur, CNRS, Laboratoire Lagrange, Boulevard de l'Observatoire,
  Nice, France}

\begin{small}
    \begin{abstract}
        We present an updated set of near-Earth asteroids with a
        Yarkovsky-related semimajor axis drift detected from the
        orbital fit to the astrometry. We find \yarko{} reliable
        detections after filtering for the signal-to-noise ratio of
        the Yarkovsky drift estimate and making sure the estimate is
        compatible with the physical properties of the analyzed
        object. Furthermore, we find a list of \marginal{} marginally
        significant detections, for which future astrometry could
        result in a Yarkovsky detection. A further outcome of the
        filtering procedure is a list of detections that we consider
        spurious because unrealistic or not explicable with the
        Yarkovsky effect. Among the smallest asteroids of our sample,
        we determined \srp{} detections of solar radiation pressure,
        in addition to the Yarkovsky effect. As the data volume
        increases in the near future, our goal is to develop methods
        to generate very long lists of asteroids with reliably
        detected Yarkovsky effect, with limited amounts of case by
        case specific adjustments. Furthermore, we discuss the
        improvements this work could bring to impact monitoring. In
        particular, we exhibit two asteroids for which the adoption of
        a non-gravitational model is needed to make reliable impact
        predictions.
    \end{abstract}
\end{small}
\maketitle

\begin{small}{\noindent\bf Keywords}: Asteroids, Yarkovsky effect,
    Non-gravitational perturbations, Orbit determination, Impact
    Monitoring
\end{small}

\section{Introduction}
Several complex phenomena cause asteroid orbital evolution to be a
difficult problem. By definition, Near-Earth Asteroids (NEAs)
experience close approaches with terrestrial planets, which are the
main source of chaos in their orbital evolution. Because of this
chaoticity, small perturbations such as non-gravitational ones can
significantly affect a NEA trajectory.

The \emph{Yarkovsky effect} is due to the recoil force undergone by a
rotating body as a consequence of its anisotropic thermal emission
\citep{vokrou:yarko,yarko_asteroids4}. The main manifestation of the
Yarkovsky effect is a secular semimajor axis drift $da/dt$, which
leads to a mean anomaly runoff that grows quadratically with
time. Typical values of this perturbation for sub-kilometre NEAs are
$da/dt\simeq 10^{-4}$--$10^{-3}$~au/My. Because of its small size, the
Yarkovsky effect can only be detected for asteroids with a well
constrained orbit.

The Yarkovsky effect is a non-gravitational perturbation generally
split into a seasonal and a diurnal component. The seasonal component
arises from the temperature variations that the heliocentric asteroid
experiences as a consequence of its orbital motion. An explicit
computation of the corresponding acceleration is not easy, even for a
spherical body, and becomes very difficult for complex shaped
bodies. On the other hand, the diurnal component is due to the lag
between the absorption of the radiation coming from the Sun, and the
corresponding re-emission in the thermal wavelength. The surface of a
rotating body illuminated by the Sun is warmed by solar radiation
during the day and cools at night. It is worth noticing that the
diurnal and the seasonal components have different consequences on the
secular semimajor axis drift. In particular, the diurnal effect
produces a positive drift for prograde rotators and a negative drift
for retrograde rotators, whereas the seasonal secular drift is always
negative. The magnitude of the diurnal effect is generally larger than
that of the seasonal effect \citep{vokrou:yarko}.

The Yarkovsky effect is key to understanding several aspects of
asteroids dynamics:
\begin{enumerate}
  \item Non-gravitational forces can be relevant for a reliable impact
    risk assessment when in the presence of deep planetary encounters
    or when having a long time horizon for the potential impact search
    \citep{im_asteroids4}. As a matter of fact, both these factors
    call for a greater consideration of the sources of orbit
    propagation uncertainty, such as non-gravitational
    perturbations. Currently, there are four known cases that required
    the inclusion of the Yarkovsky effect in term of hazard
    assessment: (101955)~Bennu
    \citep{milani:Bennu_impact,chesley:Bennu}, (99942)~Apophis
    \citep{chesley:apophis, giorgini:apophis, vokro:apophis,
      farnocchia:apophis}, (29075)~1950~DA
    \citep{giorgini:29075,farnocchia:29075}, and (410777)~2009~FD
    \citep{spoto:410777}.
  \item The semimajor axis drift produced by the Yarkovsky effect has
    sculpted the main belt for millions of years
    \citep{vokrou:families}. The Yarkovsky effect is crucial in
    understanding the aging process of asteroid families and the
    transport mechanism from the main belt to the inner Solar System
    \citep{vokrou:yarko}. The Yarkovsky effect has still not been
    measured in the main belt, thus \cite{spoto:fam_ages} used a
    calibration based on asteroid (101955)~Bennu to compute the ages
    of more than 50 families in the main belt. The improvement in the
    detection of the Yarkovsky drift for NEAs represents a new step
    forward in creating a reliable chronology of the asteroid belt.
  \item Yarkovsky detections provide constraints on asteroid physical
    properties. Two remarkable efforts in estimating the bulk density
    from the Yarkovsky drift have already been done for two
    potentially hazardous asteroids: (101955)~Bennu
    \citep{chesley:Bennu} and (29075)~1950~DA
    \citep{rozitis:29075}. Furthermore, the dependence of the
    Yarkosvky effect on the obliquity can be useful to model the spin
    axis obliquity distribution of near-Earth asteroids
    \citep{tardioli:obliquity}.
\end{enumerate}

\noindent Several efforts have been done in modelling and determining
the Yarkovsky effect on the NEA population. The first detection of the
Yarkovsky effect was predicted for the asteroid (6489)~Golevka in
\cite{vokrou:yarko} and achieved in 2003 thanks to radar ranging of
this object \citep{chesley:Golevka}. Later, the Yarkovsky effect
played a fundamental role in the attribution of four 1953 precovery
observations to the asteroid (152563)~1992~BF
\citep{vokrou:1992BF}. Moreover, \cite{chesley:Bennu} detected the
Yarkovsky effect acting on (101955)~Bennu from the astrometric
observations and from high-quality radar measurements over three
apparitions. Currently, asteroid Bennu has the best determined value
for the Yarkovsky acceleration, which also led to an estimation of its
bulk density \citep{chesley:Bennu}. More in general,
\cite{nugent:yarko} provided a list of 13 Yarkovsky detection, and
later work increased this number to 21 \citep{farnocchia:yarko}. The
most recent census is from \cite{chesley:yarko}, which identified 42
NEAs with valid Yarkovsky detection. Both \cite{farnocchia:yarko} and
\cite{chesley:yarko} flag spurious cases based on whether the detected
drift is compatible with the physical properties of the corresponding
object and the Yarkovsky mechanism. Since the number of significant
Yarkovsky detections in the NEA population is steadily increasing, we
decided to update the list.

\section{Method}
\label{sec:method}

\subsection{Force model}
\label{sub:dyn_mod}
Usually, there is not enough information on an asteroid's physical
model to directly compute the Yarkovsky acceleration through a
thermophysical model. Instead, evidence of the Yarkovsky-related drift
may be detectable from the observational dataset via orbit
determination. Indeed, a gravity-only model may not provide a
satisfactory fit to the available data. A Yarkovsky detection is more
likely when a very accurate astrometric dataset is available,
especially in case of radar measurements at multiple apparitions, or
when the observational arc is long, thus allowing the orbital drift to
become detectable. In such cases, a force model including also the
Yarkovsky acceleration could result in a better fit to the
observations.

We model the Yarkovsky perturbation with a formulation that depends on
a single dynamical parameter, to be determined in the orbital fit
together with the orbital elements. Since the secular perturbation
caused by the Yarkovsky effect is a semimajor axis drift, we use a
transverse acceleration
\begin{equation}\label{eq:modA2}
    \mathbf{a}_t = A_2g(r)\hat{\mathbf{t}}
\end{equation}
as in \cite{mardsen:comet} and \cite{farnocchia:yarko}. In equation
\eqref{eq:modA2} $A_2$ is a free parameter and $g(r)$ is a suitable
function of the heliocentric distance of the asteroid. In particular
we assume a power law
\[
    g(r) = \left(\frac{r_0}{r}\right)^d,
\]
where $r_0 = 1$ au is used as normalization factor. The exponent $d$
depends on the asteroid and is related to the asteroid's
thermophysical properties. \cite{farnocchia:yarko} show that the value
of $d$ is always between $0.5$ and $3.5$. They used $d=2$ for all
asteroids because no thermophysical data are available. In our
analysis we adopted the same values for $d$, apart from (101955)~Bennu
for which the value $d=2.25$ is assumed \citep{chesley:Bennu}.

Typical values of the Yarkovsky acceleration for a sub-kilometre NEA
are $10^{-15}$--$10^{-13}$~au/d$^2$. As a consequence, to reliably
estimate the Yarkovsky effect, the right-hand side of the equations of
motion has to include all the accelerations down to the same order of
magnitude. Our force model includes the gravitational accelerations of
the Sun, the eight planets, and the Moon based on JPL's planetary
ephemerides DE431 \citep{de431}. To ensure a more complete force
model, we also include the contributions coming from 16 massive main
belt bodies and Pluto. All the gravitational masses we used are listed
in Table~\ref{tab:ast17}. Since we compare our results with the ones
obtained by JPL, we point out that the JPL team uses the 16 most
massive main belt asteroids as estimated by \cite{de431}, which
produces a slight difference, both in the list and in the masses.

\begin{table}[h]
    \begin{center}
        \caption{Perturbing bodies included in the dynamical model in
          addition to the Sun, the planets and the Moon. They are 16
          massive main belt bodies and Pluto. The last columns shows
          the references we used for each asteroid mass.}
        \vspace{0.3cm}
        \begin{tiny}
            \begin{tabular}{lS[table-format=3.3]l}
                \hline
                \textbf{Asteroid} & {\textbf{Grav. mass}} & \textbf{Reference}\\
                &  {(km$^3$/s$^2$)} &\\
                \hline
                (1) Ceres        &   63.20 & \cite{standish:CPV}\\
                (2) Pallas       &   14.30 & \cite{standish:CPV}\\
                (3) Juno         &    1.98 & \cite{konopliv}\\
                (4) Vesta        &   17.80 & \cite{standish:CPV}\\
                (6) Hebe         &    0.93 & \cite{carry:density}\\
                (7) Iris         &    0.86 & \cite{carry:density}\\
                (10) Hygea       &    5.78 & \cite{baer:masses}\\
                (15) Eunomia     &    2.10 & \cite{carry:density}\\
                (16) Psyche      &    1.81 & \cite{carry:density}\\
                (29) Amphitrite  &    0.86 & \cite{carry:density}\\
                (52) Europa      &    1.59 & \cite{carry:density}\\
                (65) Cybele      &    0.91 & \cite{carry:density}\\
                (87) Sylvia      &    0.99 & \cite{carry:density}\\
                (88) Thisbe      &    1.02 & \cite{carry:density}\\
                (511) Davida     &    2.26 & \cite{carry:density}\\
                (704) Interamnia &    2.19 & \cite{carry:density}\\
                (134340) Pluto   &  977.00 & \cite{de431}\\
                \hline
            \end{tabular}
        \end{tiny}
        \label{tab:ast17}
    \end{center}
\end{table}

The relativity model includes the relativistic contribution of the
Sun, the planets and the Moon. In particular, we use the
Einstein-Infeld-Hoffman equations, namely the equations of the
approximate dynamics of a system of point-like masses due to their
mutual gravitational interactions, in a first order post-Newtonian
expansion, as described in \cite{eih,will:grav,moyer2003}.

\subsection{Statistical treatment of the astrometry}
The statistical treatment of the astrometry is key to a reliable orbit
determination. The differential corrections procedure provides the
asteroid's nominal orbit and its uncertainty
\citep[Chap.~5]{milani:orbdet}, which strongly depend upon the
observations accuracy and error modelling. For the computations done
for this paper, we used the debiasing and weighting scheme provided in
\cite{farnocchia:fcct}. The JPL team uses the more recent
\cite{veres:2017} weighting scheme.

The astrometric data usually can contain outliers that can affect the
solution of the orbit determination. To remove erroneous observations
from the fit we apply the outlier rejection procedure described in
\cite{carpino:outrej}.

Besides our default data treatment, we applied \emph{ad hoc}
modifications for the following cases:

\paragraph{(152563)~1992~BF.} The four 1953 precovery observations of
this NEA have been carefully re-measured in \cite{vokrou:1992BF}. We
adopt the given positions and standard deviations, the latter being
$0.5$ arcsec in right ascension and $1$ arcsec in declination.

\paragraph{2009~BD.} This object is one of the smallest near-Earth
asteroids currently known \citep{mommert:2009bd} and thus solar
radiation pressure affects its orbit. A direct detection of the
area-to-mass ratio is contained in \cite{micheli:2009BD}, which
provide high-quality astrometry from Mauna Kea and replace all the
observations from the Tzec Maun Observatory (H10) with a single
position. For these observations we set data weights based on the
uncertainties provided by \cite{micheli:2009BD} and we include both
the Yarkovsky effect and solar radiation pressure in the orbital fit
(see Section~\ref{sec:srp}).

\paragraph{2011~MD.} As well as 2009~BD, this asteroid is very small
and is among those for which we determined both the Yarkovsky effect
and solar radiation pressure. 2011~MD has been observed during the
2011 very close approach with the Earth. Despite the short arc of
three months, a very large number of optical observations of 2011~MD
were collected, precisely 1555. Following the strategy presented in
\cite{mommert:2011MD}, we relaxed the weights for the observations
collected during the close approach\footnote{Indeed, timing errors are
  more relevant for observations performed at small geocentric
  distances.} and we added the Spitzer detection (on 2014 February
11), which extends the observation arc by almost three years.

\paragraph{2015~TC$_{25}$.} Asteroid 2015~TC$_{25}$ was discovered by
the Catalina Sky Survey in October 2015, just two days before an Earth
flyby at 0.3 lunar distances. It is one of the smallest asteroids ever
discovered, about 2~m in diameter \citep{reddy:tc25}, and the 2017
astrometry permits to achieve an estimate of solar radiation
pressure. We are aware that for 2015~TC$_{25}$ the JPL team carried
out a specific study \citep{farnocchia:tc25}, which adopted \emph{ad
  hoc} weights based on observer-provided uncertainty estimates. To
handle this case, we used the same data treatment as JPL.

~\\[0.1cm]Note that it is desirable to keep the number of ``manual''
interventions on the observational data as small as possible. Indeed
we are trying to figure out how to automatize the determination of the
set of NEAs with significant and reliable Yarkovsky effect. Anyway, in
some cases a manual intervention is needed to properly handle
observational issues, \emph{e.g.} too much close observations taken
during a very close approach and affected by timing errors,
remeasurement of old observations.

\subsection{Starting sample of NEAs}\label{sub:starting}

As first sample of asteroids, we started selecting those objects in
NEODyS\footnote{The NEODyS database is available at
  \url{http://newton.dm.unipi.it}.} having a formal uncertainty on the
semimajor axis $\sigma(a)<3\cdot 10^{-9}$~au. The choice of the
threshold for $\sigma(a)$ comes from an order of magnitude estimate:
for an asteroid with diameter $1$~km the Yarkovsky drift is about
$3\cdot 10^{-10}$~au/y, thus it causes a variation of $3\cdot
10^{-9}$~au in ten years. The value of $\sigma(a)$ has to be the one
computed at the mean epoch of the observations, since it is the best
choice for the orbital fit quality. Moreover, this uncertainty
threshold corresponds to a gravity-only fit: after the Yarkovsky
coefficient is estimated the uncertainty of the semimajor axis sharply
increases because of the strong correlation between $A_2$ and the
semimajor axis.

The list of asteroids satisfying this criterion contained 519 objects
(as of February 2018). As a second step, we extracted from the JPL
database a set of \yarkojpl{} asteroids having $A_2$
determined\footnote{The JPL Small-Body Database is available at
  \url{http://ssd.jpl.nasa.gov/sbdb.cgi}.}. Among them, only $16$ were
not contained in our first list, thus we added them. Furthermore, we
considered all the reliable detections from \cite{farnocchia:yarko}
and it turned out that only 3 asteroids did not belong to any of the
previous lists, thus we added them as well to our sample.

Summarizing, we started with a sample of $519+16+3 = 538$ objects. For
each one of them we performed an orbital fit for the initial
conditions together with the Yarkovsky parameter $A_2$, without any
\emph{a priori} constraint. For few of them we also estimated solar
radiation pressure. As a result of the fit, we derived the
signal-to-noise ratio $\SNR_{A_2}$ of the $A_2$ parameter, obtaining
\snrh{} detections with $\SNR_{A_2} \geq 3$ and \snrl{} with
$\SNR_{A_2}<3$, most of which showing a negligible signal-to-noise
ratio.

\subsection{Yarkovsky expected value}
\label{sub:expected}
By means of orbit determination, we determine a transverse
acceleration directly from the astrometry. However, to claim that the
measured acceleration is caused by the Yarkovsky effect we need to
make sure that its magnitude is compatible with the physical
properties of the object and the Yarkovsky mechanism. Therefore, we
provide an expected value of the Yarkovsky-related orbital drift.

In \cite{farnocchia:yarko}, an expected value for $A_2$ is computed by
exploiting the diameter of the asteroid and scaling from the
corresponding value of (101955)~Bennu, the best determined and
reliable Yarkovsky detection. In this paper, we make use of the
Yarkovsky calibration as in \cite{spoto:fam_ages}:
\begin{equation}\label{eq:calib}
  \left(\frac{da}{dt}\right)_{\text{exp}} =
  \left(\frac{da}{dt}\right)_{\mathcal B} \cdot
  \frac{\sqrt{a_{\mathcal B}} \left(1-e_{\mathcal
      B}^2\right)}{\sqrt{a} \left(1-e^2\right)} \frac{D_{\mathcal
      B}}{D} \frac{\rho_{\mathcal B}}{\rho}
  \frac{\cos\phi}{\cos\phi_{\mathcal{B}}} \frac{1-A}{1-A_{\mathcal B}},
\end{equation}
where $D$ is the diameter of the asteroid, $\rho$ is the density,
$\phi$ is the obliquity (angle between the spin axis and the normal to
the orbit plane), and $A$ is the Bond albedo. The latter is computed
from the geometric albedo $p_v$, using $A = \frac13 p_v$
\citep{muinonen:bond}. The symbols with a ``$\mathcal{B}$'' refer to
asteroid (101955)~Bennu, and the values we assume for them are listed
in Table~\ref{tab:bennu} with their references.

\begin{table}[h]
  \begin{center}
      \caption{Values of the physical quantities for the asteroid
        (101955)~Bennu, used in equation~\eqref{eq:calib}.}
    \vspace{0.3cm}
    \begin{tiny}
      \begin{tabular}{llll}
          \hline
          \textbf{Physical quantity} & \textbf{Symbol} & \textbf{Value} & \textbf{Reference}\\
          \hline
          diameter         & $D_{\mathcal{B}}$      & $(0.492\pm 0.020)$~m       & \cite{nolan:Bennu}\\
          density          & $\rho_{\mathcal{B}}$   & $(1.26\pm 0.07)$~g/cm$^3$ & \cite{chesley:Bennu}\\
          geometric albedo & $(p_{v})_\mathcal{B}$  & $0.046\pm 0.005$       & \cite{emery:Bennu}  \\
          obliquity & $\phi_\mathcal{B}$  & $(175\pm 4)$~deg & \cite{nolan:Bennu}\\
          \hline
      \end{tabular}
    \end{tiny}
    \label{tab:bennu}
  \end{center}
\end{table}

For the diameter $D$ we use the known physical value when
available. When the asteroid's shape is not so simple to be
approximated by an ellipsoidal model, we use the dynamically
equivalent equal volume ellipsoid dimensions to compute the equivalent
diameter. In particular, this effort has been done for three
asteroids, namely (4179)~Toutatis \citep{hudson:4179},
(162421)~2000~ET$_{70}$ \citep{naidu:162421}, and
(275677)~2000~RS$_{11}$ \citep{brauer:275677}. Otherwise, when no
physical information are available, we estimate the diameter from the
absolute magnitude $H$ following the relation \citep{pravec}
\[
  D = 1329\text{ km} \cdot 10^{-H/5} \cdot \frac{1}{\sqrt{p_v}},
\]
where the geometric albedo $p_v$ is assumed to be $p_v=0.154$ if
unknown.

As shown in equation~\eqref{eq:calib}, the density is required to
estimate the strength of the Yarkovsky effect for asteroids with small
diameters. \cite{carry:density} reports a large number of asteroid
densities that we use as starting point. However, in general, density
estimates are more reliable and accurate for massive bodies and there
is a trend for a decreasing density with diameter, due to the
increasing macroporosity\footnote{It is the fraction of volume
  occupied by voids.} resulting from the cascade of collisions
suffered by the body \citep{carry:density,scheeres:astIV}.  We thus
extrapolate the density of small asteroids from the density of large
asteroids belonging to the same taxonomic class by increasing their
macroporosity to that of Bennu ($\mathcal{P}_{\mathcal{B}} = (40\pm
10)\%$, from \cite{chesley:Bennu}). Such macroporosity is typical for
(sub-)kilometre-sized asteroids, as illustrated by (25143)~Itokawa,
visited by the JAXA Hayabusa mission \citep{hayabusa}.This is a
modified version of the approach given in \cite{spoto:fam_ages}, still
using Bennu for the scaling since it has the best estimated Yarkovsky
acceleration and a comprehensive physical
characterization\footnote{Previously \cite{spoto:fam_ages} used the
  known density of (704)~Interamnia, considered to be a large asteroid
  with similar composition to Bennu, to estimate porosity of the
  latter. Recently the composition of (101955)~Bennu has been modelled
  by \cite{clark:bennu}, based on spectral observations, and it has
  been found to be closer to other large asteroids, such as
  (24)~Themis and (2)~Pallas.}. Thus the scaled density is given by
\begin{equation}\label{eq:scal_dens}
    \rho_s = (1-\mathcal{P}_{\mathcal{B}}) \rho,
\end{equation}
where the density scaling factor is $1-\mathcal{P}_{\mathcal{B}}=0.60$
and $\rho$ is the known density of the large
asteroid. Equation~\eqref{eq:scal_dens} follows from the above
assumptions and from the definition of macroporosity. We selected the
large asteroids (4)~Vesta, (10)~Hygiea, (15)~Eunomia, and
(216)~Kleopatra as representative of the taxonomic classes V, C, S, Xe
respectively.  The density of the representative asteroids and their
scaled values are listed in Table~\ref{tab:ref_ast}.

\begin{table}[h]
  \begin{center}
      \caption{Representative asteroids for some taxonomic classes:
        number/name, taxonomic type, densities as
        in~\cite{carry:density} with their uncertainties, scaled
        densities applying the factor $1-\mathcal{P}_{\mathcal{B}}$.}
    \label{tab:ref_ast}
    \vspace{0.3cm}
    \begin{tiny}
      \begin{tabular}{lccc}
          \hline
          \textbf{Asteroid} & \textbf{Tax. type} & $\rho$       & $\rho_s$\\
          &                   & (g/cm$^3$)    & (g/cm$^3$)\\
          \hline
          (4)   Vesta         &  V   &  $3.58 \pm   0.15$      & 2.15  \\
          (10)  Hygiea        &  C   &  $2.19 \pm   0.42$      & 1.31  \\
          (15)  Eunomia       &  S   &  $3.54 \pm   0.20$      & 2.12  \\
          (216) Kleopatra     &  Xe  &  $4.27 \pm   0.15$      & 2.56  \\
          \hline
      \end{tabular}
    \end{tiny}
  \end{center}
\end{table}
We used three sources of asteroid physical information: the database
of physical properties of near-Earth asteroids provided by
E.A.R.N.\footnote{\url{http://earn.dlr.de/nea/}}, the JPL Small-Body
Database, and the data provided by the WISE mission, such as diameters
and albedos \citep{wise}. It is important to point out that we have no
physical information for the large majority of the objects discussed
in this paper. For instance, for the $44\%$ of our detections with
$\SNR_{A_2}>2.5$ we have no physical data, for $62\%$ we have no
measured albedo values, and less than half of our detections can be
assigned to a taxonomic class.

\subsection{Filtering criterion}
\label{sub:calib_filter}

We use the Yarkovsky-related expected value as a filtering criterion
to understand whether the estimated orbital drift $da/dt$ is
physically consistent with the Yarkovsky effect. If the estimated
$da/dt$ is significantly larger than the maximum absolute expected
value (assuming $\cos\phi = \pm 1$), the result is inconsistent with
the Yarkovsky mechanism. We compute the indicator parameter
\begin{equation}
  \mathcal{S} = \left| \dfrac{da/dt}{(da/dt)_{\text{exp}}}\right|.
\end{equation}
Since most times there is very little to no physical information, we
need some margin on the upper threshold for $\mathcal{S}$, which
therefore should be larger than 1. We filter out the candidate
detections with $\mathcal{S} > 2$. The current maximum value allowed
for $\mathcal{S}$ is empirical, but it could be refined. In
particular, this upper threshold can be lowered when better data are
available. Improving the computation of the expected value - thus
decreasing the uncertainty of the indicator parameter $\mathcal{S}$ -
requires at least a reliable taxonomic type (for the scaling needed
for the density) and better diameters. Values of $\mathcal{S}$ greater
than the maximum threshold indicate questionable results. These
spurious detections should be investigated to find possible causes and
solutions. In general, either the $\mathcal{S}$ value is too high to
be compatible with an acceptable detection or it is barely above the
maximum threshold, in such a way that additional information would
clarify the situation and allow us to decide whether the detection is
accepted or refused. For further details see Section~\ref{sec:rej}.

We point out that values $\mathcal{S}\ll 1$ are permitted. Indeed
equation~\eqref{eq:calib} employed asteroid size and bulk density,
thus $\mathcal{S}\ll 1$ means that the orbital drift is significantly
lower than the maximum expected value. Several phenomena can lower the
Yarkovsky effect: obliquity $\phi\simeq 90^\circ$, very large or very
small thermal inertia, larger density than expected, or small rotation
angular velocity. For instance, asteroid (85774)~1998~UT$_{18}$ has a
rotation period of about 34~h, and indeed the indicator $\mathcal{S}$
is low ($\simeq 0.3$, cf.  Table~\ref{tab:reliable1}). The detections
of this kind are significant detections of a weak Yarkovsky drift. A
second class of weak Yarkovsky drifts can be defined: they are
non-detections, that is $\SNR_{A_2}<3$, but the asteroid has physical
properties that would permit a significant detection if the Yarkovsky
effect were maximized. \cite{chesley:yarko} refer to these detections
as weak detections. Despite the low $\SNR$, the result of the $A_2$
estimation can provide useful constraints on the asteroid's physical
properties.

\begin{figure}[t!]
  \centering
  \includegraphics[width=0.95\columnwidth]{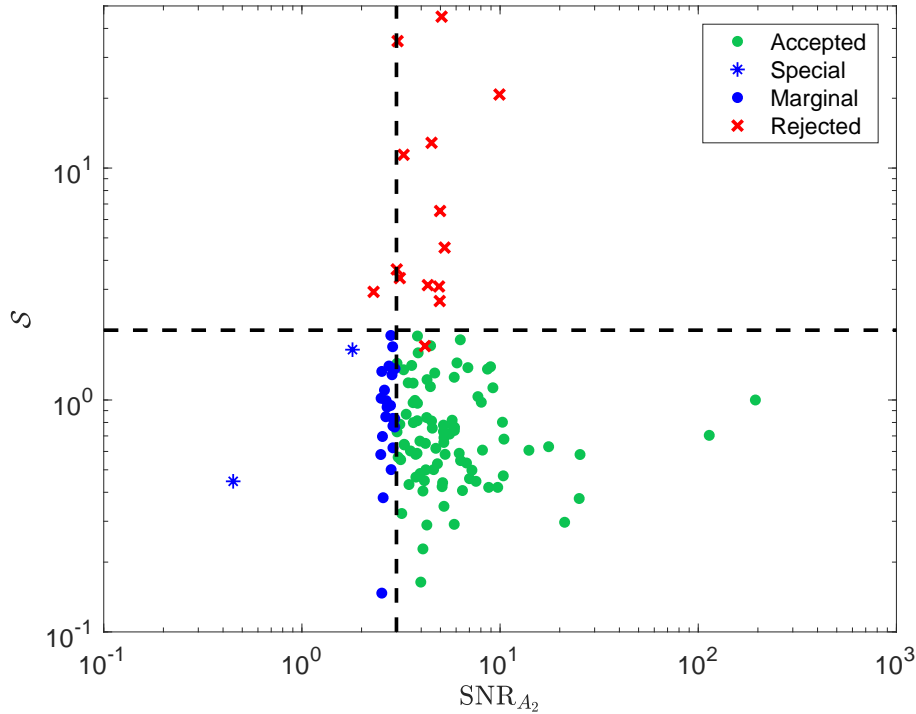}%
  \caption{Graphical representation of the partition of the detections
    set into classes on the plane $(\SNR_{A_2},\mathcal{S})$. We plot
    the accepted detections (green dots), the marginally significance
    detections (blue dots) plus the special cases (410777)~2009~FD and
    (99942)~Apophis (blue asterisk), and the rejected detections (red
    crosses).}
  \label{fig:yarko}
\end{figure}

By combining the value of the $\SNR_{A_2}$ coming from the orbital fit
with the value of the filtering parameter $\mathcal{S}$, we divided
our detections in three categories.
\begin{itemize}
  \item We consider \emph{accepted} the detections satisfying both
    $\SNR_{A_2} \geq 3$ and $\mathcal{S}\leq 2$.
  \item The category called \emph{marginal significance} includes the
    asteroids for which $2.5 < \SNR_{A_2} < 3$ and
    $\mathcal{S}\leq 2$, plus (410777)~2009~FD and (99942)~Apophis,
    both remarkable for their impact monitoring.
  \item The detections with $\SNR_{A_2}>3$ and $\mathcal{S}>2$ are
    \emph{rejected} because they have a too high value for the
    indicator parameter $\mathcal{S}$, suggesting that the detected
    $A_2$ signal is unrealistic or not explicable with the Yarkovsky
    effect (see 2003~RM in \cite{chesley:yarko}, or
    (4015)~Wilson-Harrington in Section~\ref{sec:rej}).
\end{itemize}
The results only include detections, that is we do not list the
asteroids for which we found no significant Yarkovsky signal from the
observational dataset (\nosignal~ of the initial
sample). Figure~\ref{fig:yarko} provides an overall view of our
classification. In particular, we consider the plane
$(\SNR_{A_2},\mathcal{S})$ and we mark the detections of each class
(but the rejected) with a different color:
\begin{itemize}
\item the accepted detections are indicated with a green dot;
\item the marginal significance detections are represented with a blue
  dot, except for (410777) 2009~FD and (99942)~Apophis, which are
  indicated with a blue asterisk (special cases);
\item the rejected detections are indicated with a red cross.
\end{itemize}

\section{Accepted and significant results}
\label{sec:reliable}

As explained in Section~\ref{sub:starting}, we started from the initial
sample of NEAs and we performed an orbital fit including the Yarkovsky
parameter $A_2$. Then we applied the filtering procedure, and obtained
\yarko{} detections, which are listed in Table~\ref{tab:reliable1} and
Table~\ref{tab:reliable2}. For each asteroid, we report the value of
the absolute magnitude $H$, the $A_2$ parameter along with its
uncertainty, the signal-to-noise ratio of $A_2$, the value of the
semimajor axis drift $da/dt$, the indicator parameter $\mathcal{S}$,
and the available physical data such as the geometric albedo $p_v$,
the diameter $D$, the density $\rho$, and the taxonomic class. As
explained in Section~\ref{sub:expected}, when no information on the
diameter is available, we infer one from the absolute magnitude. We
mark these cases with an asterisk ($*$) in the diameter column. When
the albedo is not directly measured but the taxonomic class is known,
we assigned the albedo according to \cite{binzel:asteroids3} and
marked the albedo value with a dagger ($\dagger$). When the albedo is
not known we assume $p_v=0.154$, and mark it with a ``d''. Note that
$0.154$ is a mean value, which has a low probability of being accurate
because of the bimodality of the albedo distribution of near-Earth
asteroids. Most asteroids are either significantly brighter or
significantly darker. Thus, when a diameter $D$ is derived from the
absolute magnitude and this default albedo, its relative uncertainty
is larger, and in turn the value of $\mathcal{S}$ (containing the
factor $1/D$) is uncertain.

\begin{table}[h]
    \setlength{\tabcolsep}{2.5pt}
    \sisetup{table-figures-uncertainty=1}
    \begin{center}
        \caption{List of Yarkovsky detections with $\SNR_{A_2}>5$ and
          with $\mathcal{S}\leq 2$. The table is sorted by
          $\SNR_{A_2}$, in decreasing order. The columns contain the
          asteroid name, the absolute magnitude $H$, the $A_2$
          parameter with its uncertainty and signal-to-noise ratio
          $\SNR_{A_2}$, the semimajor axis drift $da/dt$ with its
          uncertainty, the indicator parameter $\mathcal{S}$, the
          geometric albedo $p_v$, the diameter $D$ and the taxonomic
          class. Asteroids with no available information about the
          diamater are marked with an asterisk ($*$). Asteroids with
          albedo assigned according to \cite{binzel:asteroids3} since
          no direct estimate is available are marked with a
          ``d''. Asteroids with no albedo informations are marked with
          a dagger ($\dagger$). A remarkable detection is that of
          asteroid (480883)~2001~YE$_{4}$. This detection was not in
          the list of valid detections provided in
          \cite{chesley:yarko}, whereas in the present work it has a
          very high $\SNR_{A_2}$ value. The substantial difference
          lies in the radar observations of December 2016, confirming
          that radar is a very powerful tool in getting specifically
          high signal-to-noise ratio Yarkovsky detections.}
        \label{tab:reliable1}
        \vspace{0.3cm}
        \begin{tiny}
            \resizebox{\textwidth}{!}{
              \begin{tabular}{
                    p{2.4cm}
                    S[table-format=2.1,table-column-width=0.4cm]
                    S[table-format=4.2,table-column-width=1.5cm]@{\;$\pm$\;}
                    S[table-format=2.2,table-column-width=0.8cm]
                    S[table-format=3.2,table-column-width=0.8cm]
                    S[table-format=3.2,table-column-width=1.2cm]@{\;$\pm$\;}
                    S[table-format=2.2,table-column-width=1.2cm]
                    c
                    S[table-format=1.3,table-column-width=0.8cm]
                    S[table-format=1.3,table-column-width=0.8cm]
                    >{\centering\arraybackslash}p{0.6cm}
                  }
                  \hline
                  {\bf Asteroid} & {$H$} & \multicolumn{2}{c}{$A_2$} & {$\SNR_{A_2}$} & \multicolumn{2}{c}{$da/dt$} & {$\mathcal{S}$} & {$p_v$} & {$D$} & {Tax.}\\
                  &           & \multicolumn{2}{c}{($10^{-15}$ au/d$^2$)} &          & \multicolumn{2}{c}{($10^{-4}$ au/My)} &               &        & {(km)} & {class} \\
                  \hline

                  (101955) Bennu  & 20.6 & -46.20 & 0.24 & 194.27 & -18.98 & 0.10 & 1.0 & 0.046 & 0.492 &  B \\
(480883) 2001 YE$_{4}$  & 20.9 & -69.87 & 0.61 & 113.66 & -50.95 & 0.45 & 0.7 & 0.154$^\text{d}$ & 0.229* &  - \\
(2340) Hathor  & 20.2 & -29.94 & 1.18 & 25.32 & -17.34 & 0.69 & 0.6 & 0.6 & 0.21 &  S \\
(483656) 2005 ES$_{70}$  & 23.7 & -140.17 & 5.59 & 25.08 & -80.11 & 3.19 & 0.4 & 0.154$^\text{d}$ & 0.061* &  - \\
(152563) 1992 BF  & 19.7 & -24.85 & 1.17 & 21.17 & -11.96 & 0.56 & 0.3 & 0.287 & 0.272 &  Xc \\
2012 BB$_{124}$  & 21.1 & 71.14 & 4.05 & 17.58 & 29.42 & 1.67 & 0.6 & 0.154$^\text{d}$ & 0.201* &  - \\
(85990) 1999 JV$_{6}$  & 20.2 & -30.62 & 2.19 & 13.98 & -14.34 & 1.03 & 0.6 & 0.095 & 0.451 &  Xk \\
(437844) 1999 MN  & 21.2 & 44.56 & 4.26 & 10.46 & 41.35 & 3.95 & 0.7 & 0.154$^\text{d}$ & 0.195* &  S \\
(480808) 1994 XL$_{1}$  & 20.8 & -45.13 & 4.35 & 10.38 & -32.37 & 3.12 & 0.5 & 0.154$^\text{d}$ & 0.237* &  - \\
2007 TF$_{68}$  & 22.7 & -184.07 & 17.91 & 10.28 & -70.90 & 6.90 & 0.8 & 0.154$^\text{d}$ & 0.099* &  - \\
(1566) Icarus  & 16.3 & -3.75 & 0.39 & 9.73 & -4.85 & 0.50 & 0.4 & 0.14 & 1.44 &  S \\
(4179) Toutatis  & 15.2 & -5.95 & 0.65 & 9.20 & -2.63 & 0.29 & 1.1 & 0.13 & 2.45 &  S \\
(468468) 2004 KH$_{17}$  & 21.9 & -65.83 & 8.08 & 8.15 & -44.11 & 5.41 & 0.6 & 0.072 & 0.197 &  C \\
(138175) 2000 EE$_{104}$  & 20.3 & -106.50 & 11.89 & 8.95 & -49.37 & 5.51 & 1.4 & 0.154$^\text{d}$ & 0.297* &  - \\
(1862) Apollo  & 16.1 & -3.70 & 0.42 & 8.76 & -1.89 & 0.22 & 0.4 & 0.26 & 1.4 &  Q \\
(2062) Aten  & 17.1 & -13.18 & 1.53 & 8.64 & -5.89 & 0.68 & 1.4 & 0.2 & 1.3 &  S \\
(162004) 1991 VE  & 18.1 & 26.97 & 3.35 & 8.04 & 21.73 & 2.70 & 1.0 & 0.154$^\text{d}$ & 0.824* &  - \\
2006 TU$_{7}$  & 21.9 & 166.67 & 21.55 & 7.73 & 98.51 & 12.74 & 1.0 & 0.154$^\text{d}$ & 0.141* &  - \\
2011 PU$_{1}$  & 25.5 & -375.52 & 49.66 & 7.56 & -148.60 & 19.65 & 0.4 & 0.154$^\text{d}$ & 0.027* &  - \\
(6489) Golevka  & 19.0 & -12.04 & 1.67 & 7.21 & -5.10 & 0.71 & 0.5 & 0.151 & 0.53 &  Q \\
2011 EP$_{51}$  & 25.3 & -359.14 & 51.20 & 7.01 & -185.09 & 26.39 & 0.5 & 0.154$^\text{d}$ & 0.029* &  - \\
(33342) 1998 WT$_{24}$  & 17.8 & -27.87 & 4.05 & 6.88 & -16.91 & 2.46 & 1.4 & 0.75 & 0.415 &  Xe \\
(3361) Orpheus  & 19.2 & 18.27 & 2.70 & 6.77 & 7.88 & 1.16 & 0.5 & 0.357 & 0.348 &  Q \\
(364136) 2006 CJ  & 20.1 & -29.16 & 4.52 & 6.46 & -34.99 & 5.42 & 0.4 & 0.154$^\text{d}$ & 0.317* &  - \\
(499998) 2011 PT  & 24.0 & -234.96 & 37.16 & 6.32 & -91.30 & 14.44 & 0.5 & 0.154$^\text{d}$ & 0.053* &  - \\
(138404) 2000 HA$_{24}$  & 19.1 & 45.05 & 7.15 & 6.30 & 19.95 & 3.17 & 1.8 & 0.154$^\text{d}$ & 0.517* &  S \\
2006 CT  & 22.3 & -112.43 & 18.09 & 6.22 & -48.14 & 7.74 & 0.6 & 0.154$^\text{d}$ & 0.119* &  - \\
(3908) Nyx  & 17.3 & 25.45 & 4.20 & 6.06 & 9.86 & 1.63 & 1.4 & 0.23 & 1 &  V \\
(363599) 2004 FG$_{11}$  & 21.0 & -59.90 & 10.17 & 5.89 & -42.39 & 7.20 & 0.8 & 0.306 & 0.152 &  V \\
1999 UQ  & 21.7 & -110.45 & 18.77 & 5.88 & -44.85 & 7.62 & 0.7 & 0.154$^\text{d}$ & 0.152* &  - \\
2003 YL$_{118}$  & 21.6 & -172.62 & 29.42 & 5.87 & -90.31 & 15.39 & 1.3 & 0.154$^\text{d}$ & 0.165* &  - \\
(154590) 2003 MA$_{3}$  & 21.6 & -77.01 & 13.11 & 5.87 & -37.11 & 6.32 & 0.3 & 0.530 & 0.086 &  - \\
2005 EY$_{169}$  & 22.1 & -137.02 & 23.70 & 5.78 & -53.80 & 9.30 & 0.8 & 0.154$^\text{d}$ & 0.128* &  - \\
(10302) 1989 ML  & 19.4 & 74.98 & 13.09 & 5.73 & 28.76 & 5.02 & 0.8 & 0.51 & 0.248 &  - \\
2000 PN$_{8}$  & 22.1 & 123.75 & 22.26 & 5.56 & 49.28 & 8.87 & 0.7 & 0.154$^\text{d}$ & 0.131* &  - \\
(506590) 2005 XB$_{1}$  & 21.9 & 92.68 & 17.54 & 5.28 & 44.88 & 8.49 & 0.6 & 0.154$^\text{d}$ & 0.143* &  - \\
(350462) 1998 KG$_{3}$  & 22.1 & -61.35 & 11.79 & 5.21 & -24.52 & 4.71 & 0.3 & 0.154$^\text{d}$ & 0.129* &  - \\
(216523) 2001 HY$_{7}$  & 20.5 & 58.55 & 11.23 & 5.21 & 31.33 & 6.01 & 0.7 & 0.154$^\text{d}$ & 0.267* &  - \\
(363505) 2003 UC$_{20}$  & 18.3 & -7.48 & 1.44 & 5.20 & -4.05 & 0.78 & 0.7 & 0.028 & 1.9 &  C \\
(99907) 1989 VA  & 17.9 & 16.51 & 3.19 & 5.18 & 12.71 & 2.46 & 0.7 & 0.24 & 0.55 &  S \\
(66400) 1999 LT$_{7}$  & 19.3 & -43.09 & 8.33 & 5.18 & -29.44 & 5.69 & 0.8 & 0.182 & 0.411 &  - \\
(377097) 2002 WQ$_{4}$  & 19.5 & -23.66 & 4.61 & 5.13 & -10.37 & 2.02 & 0.4 & 0.154$^\text{d}$ & 0.423* &  - \\
2000 CK$_{59}$  & 24.2 & -192.46 & 37.77 & 5.10 & -74.48 & 14.62 & 0.4 & 0.154$^\text{d}$ & 0.05* &  - \\

                  \hline
              \end{tabular}}
        \end{tiny}
    \end{center}
\end{table}

\begin{table}[h]
    \setlength{\tabcolsep}{2.5pt}
    \sisetup{table-figures-uncertainty=1}
    \begin{center}
        \caption{List of Yarkovsky detections with $3\leq\SNR_{A_2}<5$
          and with $\mathcal{S}\leq 2$. The table is sorted by
          $\SNR_{A_2}$, in decreasing order. Columns and symbols are
          the same as in Table~\ref{tab:reliable1}.}
        \label{tab:reliable2}
        \vspace{0.3cm}
        \begin{tiny}
            \resizebox{\textwidth}{!}{
              \begin{tabular}{
                    p{2.4cm}
                    S[table-format=2.1,table-column-width=0.54cm]
                    S[table-format=4.2,table-column-width=1.5cm]@{$\pm$}
                    S[table-format=2.2,table-column-width=0.8cm]
                    S[table-format=1.2,table-column-width=0.8cm]
                    S[table-format=3.2,table-column-width=1.2cm]@{$\pm$}
                    S[table-format=2.2,table-column-width=1.2cm]
                    c
                    S[table-format=1.3,table-column-width=0.8cm]
                    S[table-format=1.3,table-column-width=0.8cm]
                    >{\centering\arraybackslash}p{0.6cm}
                  }
                  \hline
                  {\bf Asteroid} & {$H$} & \multicolumn{2}{c}{$A_2$} & {$\SNR_{A_2}$} & \multicolumn{2}{c}{$da/dt$} & {$\mathcal{S}$} & {$p_v$} & {$D$} & {Tax.}\\
                  &           & \multicolumn{2}{c}{($10^{-15}$ au/d$^2$)} &          & \multicolumn{2}{c}{($10^{-4}$ au/My)} &               &        & {(km)} & {class} \\
                  \hline

                  (29075) 1950 DA  & 17.1 & -6.03 & 1.25 & 4.83 & -2.65 & 0.55 & 0.5 & 0.07 & 2 &  - \\
(162117) 1998 SD$_{15}$  & 19.1 & -15.55 & 3.28 & 4.74 & -7.76 & 1.64 & 0.6 & 0.154$^\text{d}$ & 0.51* &  S \\
2001 BB$_{16}$  & 23.0 & 345.54 & 73.84 & 4.68 & 163.59 & 34.96 & 1.3 & 0.154$^\text{d}$ & 0.086* &  - \\
(138852) 2000 WN$_{10}$  & 20.1 & 36.04 & 7.80 & 4.62 & 16.80 & 3.64 & 0.5 & 0.154$^\text{d}$ & 0.316* &  - \\
(455176) 1999 VF$_{22}$  & 20.7 & -69.25 & 15.23 & 4.55 & -56.46 & 12.41 & 0.8 & 0.154$^\text{d}$ & 0.248* &  - \\
(399308) 1993 GD  & 20.6 & 102.49 & 22.73 & 4.51 & 43.94 & 9.75 & 0.8 & 0.3 & 0.18 &  - \\
(7336) Saunders  & 18.8 & 39.34 & 8.82 & 4.46 & 14.29 & 3.20 & 1.7 & 0.18$^\dagger$ & 0.553* &  S \\
(1685) Toro  & 14.3 & -3.76 & 0.84 & 4.45 & -1.68 & 0.38 & 1.1 & 0.26 & 3.75 &  S \\
(4034) Vishnu  & 18.3 & -66.24 & 15.48 & 4.28 & -34.03 & 7.95 & 1.2 & 0.52 & 0.42 &  - \\
(85774) 1998 UT$_{18}$  & 19.1 & -6.64 & 1.55 & 4.27 & -2.67 & 0.62 & 0.3 & 0.042 & 0.939 &  C \\
(310442) 2000 CH$_{59}$  & 19.8 & 52.16 & 12.25 & 4.26 & 29.04 & 6.82 & 0.8 & 0.154$^\text{d}$ & 0.366* &  - \\
(2100) Ra-Shalom  & 16.2 & -4.65 & 1.10 & 4.22 & -2.67 & 0.63 & 0.5 & 0.14 & 2.24 &  C \\
(326354) 2000 SJ$_{344}$  & 22.8 & -158.81 & 37.77 & 4.20 & -65.15 & 15.49 & 0.7 & 0.154$^\text{d}$ & 0.093* &  - \\
(481442) 2006 WO$_{3}$  & 21.6 & -62.26 & 15.00 & 4.15 & -36.97 & 8.90 & 0.4 & 0.154$^\text{d}$ & 0.164* &  - \\
(306383) 1993 VD  & 21.4 & -29.85 & 7.32 & 4.08 & -19.46 & 4.77 & 0.2 & 0.154$^\text{d}$ & 0.174* &  - \\
(441987) 2010 NY$_{65}$  & 21.5 & -37.87 & 9.28 & 4.08 & -18.65 & 4.57 & 0.4 & 0.071 & 0.228 &  C \\
2008 CE$_{119}$  & 25.6 & -143.47 & 36.04 & 3.98 & -57.16 & 14.36 & 0.2 & 0.154$^\text{d}$ & 0.026* &  - \\
(85953) 1999 FK$_{21}$  & 18.1 & -9.85 & 2.49 & 3.95 & -9.63 & 2.44 & 0.5 & 0.32 & 0.59 &  S \\
(348306) 2005 AY$_{28}$  & 21.6 & -91.12 & 23.12 & 3.94 & -61.39 & 15.58 & 0.7 & 0.154$^\text{d}$ & 0.166* &  - \\
(65679) 1989 UQ  & 19.4 & -37.59 & 9.74 & 3.86 & -17.95 & 4.65 & 1.6 & 0.033 & 0.918 &  C \\
1995 CR  & 21.7 & -85.94 & 22.44 & 3.83 & -155.89 & 40.71 & 1.0 & 0.18$^\dagger$ & 0.143* &  S \\
(232691) 2004 AR$_{1}$  & 19.8 & -116.25 & 30.33 & 3.83 & -50.45 & 13.16 & 1.9 & 0.154$^\text{d}$ & 0.369* &  - \\
(265482) 2005 EE  & 21.2 & 93.97 & 24.62 & 3.82 & 42.07 & 11.02 & 0.8 & 0.154$^\text{d}$ & 0.197* &  - \\
(136818) Selqet  & 19.0 & 24.44 & 6.42 & 3.81 & 12.18 & 3.20 & 0.6 & 0.15$^\dagger$ & 0.548* &  X \\
(425755) 2011 CP$_{4}$  & 21.1 & 52.62 & 13.99 & 3.76 & 96.46 & 25.65 & 0.5 & 0.154$^\text{d}$ & 0.201* &  - \\
(192559) 1998 VO  & 20.4 & -33.01 & 8.81 & 3.75 & -14.25 & 3.80 & 0.6 & 0.28 & 0.216 &  S \\
(163023) 2001 XU$_{1}$  & 19.2 & 47.27 & 12.70 & 3.72 & 32.04 & 8.61 & 1.0 & 0.154$^\text{d}$ & 0.479* &  - \\
(5604) 1992 FE  & 17.2 & -24.03 & 6.61 & 3.64 & -12.68 & 3.49 & 1.2 & 0.48 & 0.55 &  V \\
(397326) 2006 TC$_{1}$  & 19.0 & 33.65 & 9.23 & 3.65 & 12.68 & 3.48 & 0.8 & 0.154$^\text{d}$ & 0.54* &  - \\
(208023) 1999 AQ$_{10}$  & 20.4 & -44.41 & 12.21 & 3.64 & -20.66 & 5.68 & 1.0 & 0.154$^\text{d}$ & 0.281* &  S \\
(437841) 1998 HD$_{14}$  & 20.9 & -87.22 & 24.35 & 3.58 & -41.83 & 11.68 & 1.4 & 0.18$^\dagger$ & 0.205* &  Q \\
(413260) 2003 TL$_{4}$  & 19.5 & -36.09 & 10.21 & 3.53 & -20.36 & 5.76 & 0.6 & 0.22 & 0.38 &  - \\
(4581) Asclepius  & 20.7 & -40.76 & 11.76 & 3.47 & -19.62 & 5.66 & 0.4 & 0.154$^\text{d}$ & 0.241* &  - \\
(136582) 1992 BA  & 19.9 & -54.38 & 16.17 & 3.36 & -20.03 & 5.96 & 0.9 & 0.154$^\text{d}$ & 0.363* &  - \\
(467351) 2003 KO$_{2}$  & 20.4 & 97.34 & 28.27 & 3.44 & 65.59 & 19.05 & 1.2 & 0.154$^\text{d}$ & 0.277* &  - \\
(7341) 1991 VK  & 16.8 & -6.04 & 1.84 & 3.29 & -2.54 & 0.77 & 0.6 & 0.18$^\dagger$ & 1.344* &  S \\
(256004) 2006 UP  & 23.0 & -174.21 & 53.10 & 3.28 & -64.61 & 19.69 & 0.6 & 0.154$^\text{d}$ & 0.084* &  - \\
(450300) 2004 QD$_{14}$  & 20.6 & -116.65 & 35.73 & 3.26 & -57.61 & 17.65 & 1.4 & 0.154$^\text{d}$ & 0.263* &  - \\
(477719) 2010 SG$_{15}$  & 25.2 & -237.31 & 74.49 & 3.19 & -90.57 & 28.43 & 0.3 & 0.154$^\text{d}$ & 0.031* &  - \\
(37655) Illapa  & 17.8 & -13.41 & 4.26 & 3.15 & -10.81 & 3.43 & 0.6 & 0.154$^\text{d}$ & 0.938* &  - \\
(267759) 2003 MC$_{7}$  & 18.7 & -29.24 & 9.36 & 3.12 & -10.97 & 3.51 & 0.8 & 0.154$^\text{d}$ & 0.611* &  - \\
(310842) 2003 AK$_{18}$  & 19.7 & -33.50 & 10.94 & 3.06 & -17.83 & 5.82 & 0.6 & 0.154$^\text{d}$ & 0.385* &  - \\
(162783) 2000 YJ$_{11}$  & 20.6 & -127.26 & 42.13 & 3.02 & -49.85 & 16.50 & 1.4 & 0.154$^\text{d}$ & 0.257* &  - \\
(152671) 1998 HL$_{3}$  & 20.1 & -55.64 & 18.40 & 3.02 & -25.68 & 8.49 & 0.7 & 0.2 & 0.298 &  - \\

                  \hline
              \end{tabular}}
        \end{tiny}
    \end{center}
\end{table}

Some of the asteroids included in Table~\ref{tab:reliable1} and
Table~\ref{tab:reliable2} deserves dedicated comments.

\paragraph{(1566)~Icarus.} It is known that the 1968 observations of
(1566)~Icarus are affected by large timing errors. A possible solution
to this problem is to include timing errors in the observations
uncertainty possibly even removing the systematic timing errors. A
possible alternative is to properly treat the correlation between the
right ascension and the declination. This operation will be made
easier after the adoption of the new Astrometric Data Exchange
Standard
(ADES\footnote{\url{http://minorplanetcenter.net/iau/info/IAU2015_ADES.pdf}.}). For
now, it is possible to adapt the weighting scheme to underweight the
observations during the 1968 close approach: this was done by the JPL
team, not by the Pisa one. By comparing the result of the two groups,
and also with the one in \cite{greenberg:icarus}, we can claim that
the detection of the Yarkovsky effect is confirmed, even if there is a
significant difference between the standard deviations (see
Section~\ref{sec:compare_JPL}), which is explained by the different
weighting scheme.

\paragraph{(3908)~Nyx.} Asteroid (3908)~Nyx is classified as V-type,
but it has many properties inconsistent with (4)~Vesta. Thus the
density scaling is not performed using the density of Vesta as for the
other V-type asteroids. Asteroid (5381)~Sekhmet is a V-type with a
diameter which is comparable to the one of Nyx. The density of Sekhmet
is $(1.30\pm 0.65)$~g/cm$^3$ \citep{carry:density}, compatible with
the estimate in \cite{farnocchia:nyx}, and we assume this value also
for Nyx.

\paragraph{(4179)~Toutatis} This asteroid shows a significant
  Yarkovsky detection with $\SNR_{A_2} > 3$ and $\mathcal{S} < 2$ and
  is therefore included in the list of accepted detections. However,
  we point out that this detection is subject to substantial
  uncertainties beyond the formal ones resulting from the fit to the
  optical and radar astrometry. The large aphelion of the orbit of
  Toutatis and the small magnitude of the Yarkovsky perturbation make
  the $A_2$ estimate sensitive to the set of main belt perturbers
  included in the force model and the uncertainty in their masses.

\section{Rejected results}
\label{sec:rej}

In this section we consider the significant detections that we rated
as spurious, \emph{i.e.} for which we obtained a Yarkovsky detection
greater than one would reasonably expect from the Yarkovsky
effect. Despite the signal-to-noise ratio is less than $3$, we also
add (4015)~Wilson-Harrington to this category as a ``special'' case,
as explained below.

Reasons for refusing a detection can be the following:
\begin{itemize}
  \item Dynamical model problems can occur in few cases, such as
    (4015) Wilson-Harrington.
  \item Sometimes the results are strongly dependent on few
    observations, typically old isolated observations, which are
    separated by a long time interval from the bulk of the dataset. In
    these cases we usually reject the detection, unless the precovery
    has been carefully remeasured, as for (152563)~1992~BF.
  \item Solutions with Yarkovsky affected by observational data of
    questionable reliability.
\end{itemize}
The list of all the rejected detections contains \rejected{} cases,
see Table~\ref{tab:refused}. Below we provide dedicated comments for
each rejected detection.

\begin{table}[h]
    \setlength{\tabcolsep}{2.5pt}
    \sisetup{table-figures-uncertainty=1}
    \begin{center}
        \caption{List of rejected Yarkovsky detections. The table is
          sorted by $\SNR_{A_2}$, in decreasing order. The columns are
          the same as in Table~\ref{tab:reliable1}, but the one
          showing $da/dt$.}
        \label{tab:refused}
        \vspace{0.3cm}
        \begin{tiny}
              \begin{tabular}{
                    p{3.5cm}
                    S[table-format=2.1,table-column-width=0.5cm]
                    S[table-format=5.2,table-column-width=1.5cm]@{$\pm$}
                    S[table-format=3.2,table-column-width=1.2cm]
                    S[table-format=1.2,table-column-width=1.0cm]
                    c
                    S[table-format=1.3,table-column-width=1.0cm]
                    S[table-format=1.3,table-column-width=1.0cm]
                    >{\centering\arraybackslash}p{0.6cm}
                  }
                  \hline
                  {\bf Asteroid} & {$H$} & \multicolumn{2}{c}{$A_2$} & {$\SNR_{A_2}$} & {$\mathcal{S}$} & {$p_v$} & {$D$} & {Tax.}\\
                  &           & \multicolumn{2}{c}{($10^{-15}$ au/d$^2$)} &      &    &   & {(km)} & {class} \\
                  \hline

                  (4015) Wilson-Harrington  & 16.0 & -16.48 & 7.16 & 2.30 & 2.9 & 0.046 & 3.821 &  C \\
\hline
(260141) 2004 QT$_{24}$  & 18.3 & 530.90 & 53.46 & 9.93 & 20.8 & 0.42 & 0.454* &  S \\
(350751) 2002 AW  & 20.7 & -579.13 & 116.36 & 4.98 & 6.5 & 0.154$^\text{d}$ & 0.243* &  B \\
(39565) 1992 SL  & 18.4 & -100.52 & 20.41 & 4.92 & 3.1 & 0.154$^\text{d}$ & 0.698* &  - \\
(4486) Mithra  & 15.4 & -83.37 & 18.47 & 4.51 & 12.9 & 0.297 & 1.849 &  V \\
(474158) 1999 FA  & 20.6 & -93.01 & 22.25 & 4.18 & 1.7 & 0.18$^\dagger$ & 0.233* &  S \\
(162421) 2000 ET$_{70}$  & 18.0 & -33.73 & 10.80 & 3.12 & 3.4 & 0.15$^\dagger$ & 2.26 &  - \\
\hline
(308635) 2005 YU$_{55}$  & 21.6 & -317.23 & 60.43 & 5.25 & 4.5 & 0.065 & 0.306 &  C \\
(139359) 2001 ME$_{1}$  & 16.6 & -307.68 & 60.74 & 5.07 & 45.0 & 0.04 & 3.15* &  C \\
(433) Eros  & 10.8 & -1.96 & 0.40 & 4.96 & 2.7 & 0.25 & 16.84 &  S \\
(175706) 1996 FG$_{3}$  & 18.3 & -55.77 & 12.90 & 4.32 & 3.1 & 0.072 & 1.196 &  C \\
2010 KP$_{10}$  & 23.4 & 2981.28 & 915.15 & 3.26 & 11.4 & 0.101 & 0.087 &  - \\
(142561) 2002 TX$_{68}$  & 18.1 & -466.98 & 153.85 & 3.04 & 35.3 & 0.154$^\text{d}$ & 0.801* &  Xe \\
(192563) 1998 WZ$_{6}$  & 17.3 & -54.76 & 18.17 & 3.01 & 3.7 & 0.30 & 0.8 &  V \\

                  \hline
              \end{tabular}
        \end{tiny}
    \end{center}
\end{table}

\paragraph{(4015)~Wilson-Harrington.} This asteroid was initially
discovered in 1949 as a comet at the Palomar Sky Survey. It was named
107P/Wilson-Harrington, but then it was lost. Thirty years later the
asteroid 1979~VA was discovered and, after the 1988 apparition, it was
numbered as (4015)~1979~VA. On August 13, 1992, the IAU circular 5585
\citep{4015:iau} reported that the asteroid (4015)~1979~VA and the
comet 107P/Wilson-Harrington were indeed the same object. Furthermore,
no cometary activity was noted during the well-observed 1979-80
apparition, confirming that it is actually an extinct comet. The
detection of the Yarkovsky parameter is indeed significant, but the
$\mathcal{S}$ value indicates a value of the non-gravitational
acceleration that is too large than the one expected from the
Yarkovsky effect. Since the observed arc contains the time span of
cometary activity, the most likely interpretation is that the large
detected transverse acceleration is caused by the out-gassing rather
than the Yarkovsky effect. Furthermore, a dynamical model assuming a
constant value for $A_2$, as it is the one we employed, is not enough
representative of the real orbital dynamics, given that the cometary
activity has ceased. Thus in this case we consider that a
non-gravitational effect has been detected, but not Yarkovsky.

In some cases, a spurious detections is due to poor optical
astrometry, often affecting isolated old observations. In this case we
rejected the detection, and a remeasurement of these old observations
would be desirable to clarify the situation: the Yarkovsky signal
could significantly increase as well as disappear. This is the case
for the asteroids listed below.

\paragraph{(260141)~2004~QT$_{24}$.} The detected signal is strongly
dependent on four observations in $1993$ and $1998$ from Siding Spring
Observatory DSS.

\paragraph{(350751)~2002~AW.} This asteroid has two isolated
observations in $1991$ from Palomar Mountain-DSS.

\paragraph{(39565)~1992~SL.} This asteroid has one isolated
observation in $1950$ from Palomar Mountain.

\paragraph{(4486)~Mithra.} This asteroid has a signal which is
strongly dependent on a single isolated observation in $1974$ from
Crimea-Nauchnij (MPC code 095).

\paragraph{(474158)~1999~FA.} This object has one isolated observation
in 1978, from Siding Spring Observatory. In agreement with
\cite{farnocchia:yarko}, we consider that the 1978 observation would
need to be remeasured before accepting the Yarkovsky detection for
(474158)~1999~FA.

\paragraph{(162421)~2000~ET$_{70}$.} This asteroid has two isolated
observations in $1977$ from European Southern Observatory, La Silla
DSS.

~\\[0.1cm]There are detections which have to be rated as
spurious, because $\mathcal{S}$ indicates a Yarkovsky drift which is
way larger than expected, despite the fact that the signal-to-noise
ratio of the $A_2$ parameter is greater than $3$. This holds for
2010~KP$_{10}$, (308635)~2005~YU$_{55}$, (139359)~2001~ME$_{1}$,
(142561)~2002~TX$_{68}$, and (192563)~1998~WZ$_{6}$. To confirm the
reliability of our filtering criterion, we carefully checked each of
these spurious detections. They show problematic astrometry, which
resulted in a incorrect determination of the Yarkovsky effect.

A separate comment holds for (175706)~1996~FG$_3$, since it is a
binary asteroid \citep{1996FG3}. The signal found for the Yarkovsky
detection is likely due to the astrometric data treatment, as
confirmed by the fact that it disappears when the weighting scheme
proposed in \cite{veres:2017} is applied. Once we have a Yarkovsky
detection for this object it will be possible to compare with the
Yarkovsky theory for binary asteroids as described by
\cite{vokro:yarko_binary}.

A remarkable case is (433)~Eros. This asteroid shows a significant
value for the Yarkovsky effect, but a very high value for the
indicator parameter. Moreover, the obliquity of Eros is known to be
$\simeq 89^{\circ}$ \citep{yeomans:eros}, therefore we would expect a
value for $\mathcal{S}$ much less than $1$. Thus this detection is
spurious, likely caused by historical data dating back to 1893 for
which it is challenging to come up with a reliable statistical
treatment.

\section{Marginal significance}
\label{sec:weak_det}

We now consider the marginal significance class, containing the
detections for which $2.5<\SNR_{A_2}<3$ and $\mathcal{S}\leq 2$. These
detections are physically meaningful since they satisfy the filtering
on $\mathcal{S}$, but the signal to noise for the $A_2$ parameter as
determined from the observations is not enough for a reliable
detection. In addition, as mentioned before, we include two special
cases in this category, namely (410777)~2009~FD and
(99942)~Apophis. These two objects show acceptable values of the
indicator $\mathcal{S}$ but the signal-to-noise ratio of the $A_2$
parameter is very low (cf. Table~\ref{tab:weak}). Nevertheless, we
decided to keep them because the Yarkovsky drift plays a fundamental
role for its impact predictions (see the introduction). In this way we
grouped \marginal{} detections in this class, which are listed in
Table~\ref{tab:weak}.

\paragraph{(99942)~Apophis.} Similarly to Toutatis, also Apophis has a
complex rotation, as shown in \cite{pravec:apophis,
  vokro:apophis}. However, the Yarkovsky effect is not significantly
weakened by the tumbling state. \cite{vokro:apophis} used the
available rotation state, shape, size and thermophysical model of
Apophis to predict the Yarkovsky semimajor axis drift. The drift
obtained by fitting the astrometric data is compatible with the model
prediction. We obtained $da/dt = (-24.50\pm 13.58)\cdot
10^{-4}$~au/Myr for the fitted value, which is completely consistent
with \cite{vokro:apophis}. There is no question that the Yarkovsky
effect has to be taken into account for Apophis to predict future
motion, especially for impact hazard assessment
\citep{chesley:apophis, giorgini:apophis, farnocchia:apophis}.

\paragraph{(410777)~2009~FD.} The Yarkovsky effect found is below the
significance level, and nevertheless it has to be taken into account
for long-term impact monitoring purposes \citep{spoto:410777}.

~\\[0.1cm]Maintaining a list of marginal significance detections is
useful because they are candidates for future detections as
observational data improves and increases.

\begin{table}[h]
    \setlength{\tabcolsep}{2.5pt}
    \sisetup{table-figures-uncertainty=1}
    \begin{center}
        \caption{List of marginal significance detections, which means
          $2.5<\SNR_{A_2}<3$ and $\mathcal{S}\leq 2$. The table is
          sorted by $\SNR_{A_2}$, in decreasing order (apart from the
          two special cases at the top). Columns and symbols are the
          same as in Table~\ref{tab:reliable1}.}
        \label{tab:weak}
        \vspace{0.3cm}
        \begin{tiny}
            \resizebox{\textwidth}{!}{
              \begin{tabular}{
                    p{2.8cm}
                    S[table-format=2.1,table-column-width=0.5cm]
                    S[table-format=4.2,table-column-width=1.5cm]@{$\pm$}
                    S[table-format=2.2,table-column-width=0.8cm]
                    S[table-format=3.1,table-column-width=0.8cm]
                    S[table-format=3.2,table-column-width=1.2cm]@{$\pm$}
                    S[table-format=2.2,table-column-width=1.2cm]
                    c
                    S[table-format=1.3,table-column-width=0.8cm]
                    S[table-format=1.3,table-column-width=0.8cm]
                    >{\centering\arraybackslash}p{0.6cm}
                  }
                  \hline
                  {\bf Asteroid} & {$H$} & \multicolumn{2}{c}{$A_2$} & {$\text{SNR}_{A_2}$} & \multicolumn{2}{c}{$da/dt$} & {$\mathcal{S}$} & {$p_v$} & {$D$} & {Tax.}\\
                  &           & \multicolumn{2}{c}{($10^{-15}$ au/d$^2$)} &          & \multicolumn{2}{c}{($10^{-4}$ au/My)} &               &        & {(km)} & {class}   \\
                  \hline

                  (99942) Apophis  & 18.9 & -53.39 & 29.60 & 1.80 & -24.50 & 13.58 & 1.6 & 0.30 & 0.375 &  S \\
(410777) 2009 FD  & 22.1 & 21.49 & 47.40 & 0.45 & 11.18 & 24.66 & 0.4 & 0.01 & 0.472 &  - \\
\hline
(162080) 1998 DG$_{16}$  & 19.8 & -37.93 & 12.84 & 2.95 & -19.51 & 6.61 & 1.4 & 0.035 & 0.777 &  C \\
(85770) 1998 UP$_{1}$  & 20.4 & -34.77 & 11.84 & 2.94 & -16.77 & 5.71 & 0.8 & 0.154$^\text{d}$ & 0.282* &  S \\
(162142) 1998 VR  & 18.7 & 17.59 & 5.98 & 2.94 & 8.88 & 3.02 & 0.8 & 0.18$^\dagger$ & 0.6 &  S \\
2002 LY$_{1}$  & 22.4 & -166.14 & 57.73 & 2.88 & -84.31 & 29.30 & 0.8 & 0.154$^\text{d}$ & 0.114* &  - \\
(474163) 1999 SO$_{5}$  & 20.9 & -79.89 & 27.78 & 2.88 & -32.69 & 11.37 & 0.8 & 0.154$^\text{d}$ & 0.22* &  - \\
(242191) 2003 NZ$_{6}$  & 19.0 & 38.23 & 13.29 & 2.88 & 24.07 & 8.37 & 0.6 & 0.334 & 0.370 &  - \\
(215588) 2003 HF$_{2}$  & 19.4 & -79.07 & 27.60 & 2.87 & -58.53 & 20.43 & 1.7 & 0.118 & 0.488 &  - \\
(162181) 1999 LF$_{6}$  & 18.2 & -22.41 & 7.86 & 2.85 & -8.70 & 3.05 & 1.3 & 0.175 & 0.729 &  S \\
(164207) 2004 GU$_{9}$  & 21.1 & -69.93 & 24.83 & 2.82 & -30.24 & 10.74 & 0.5 & 0.219 & 0.163 &  - \\
2001 QC$_{34}$  & 20.1 & -73.87 & 26.33 & 2.81 & -30.61 & 10.91 & 1.9 & 0.154$^\text{d}$ & 0.329* &  Q \\
(283457) 2001 MQ$_{3}$  & 18.9 & -38.45 & 13.73 & 2.80 & -13.80 & 4.93 & 0.9 & 0.154$^\text{d}$ & 0.56* &  - \\
2007 PB$_{8}$  & 21.2 & -160.83 & 58.50 & 2.75 & -90.77 & 33.01 & 1.4 & 0.154$^\text{d}$ & 0.198* &  - \\
(230111) 2001 BE$_{10}$  & 19.2 & -28.81 & 10.72 & 2.69 & -15.61 & 5.81 & 0.9 & 0.253 & 0.4 &  S \\
1999 SK$_{10}$  & 19.7 & -45.84 & 17.24 & 2.66 & -18.21 & 6.85 & 1.0 & 0.346 & 0.259 &  S \\
(338292) 2002 UA$_{31}$  & 19.0 & -35.78 & 13.48 & 2.65 & -22.29 & 8.40 & 0.8 & 0.154$^\text{d}$ & 0.538* &  - \\
(334412) 2002 EZ$_{2}$  & 20.1 & -119.39 & 45.75 & 2.61 & -45.46 & 17.42 & 1.1 & 0.40 & 0.21 &  - \\
(376879) 2001 WW$_{1}$  & 22.0 & -63.88 & 24.83 & 2.57 & -25.03 & 9.73 & 0.4 & 0.154$^\text{d}$ & 0.135* &  - \\
(416151) 2002 RQ$_{25}$  & 20.6 & 55.10 & 21.64 & 2.55 & 24.49 & 9.62 & 0.7 & 0.154$^\text{d}$ & 0.262* &  C \\
(503941) 2003 UV$_{11}$  & 19.5 & 6.66 & 2.63 & 2.53 & 5.62 & 2.22 & 0.1 & 0.376 & 0.26 &  Q \\
(471240) 2011 BT$_{15}$  & 21.7 & -196.07 & 77.53 & 2.53 & -80.55 & 31.85 & 1.3 & 0.154$^\text{d}$ & 0.154* &  - \\
1994 CJ$_{1}$  & 21.5 & -138.42 & 55.12 & 2.51 & -53.87 & 21.45 & 1.0 & 0.154$^\text{d}$ & 0.167* &  - \\
(54509) YORP  & 22.6 & -74.61 & 29.88 & 2.50 & -33.45 & 13.40 & 0.6 & 0.154$^\text{d}$ & 0.1 &  S \\

                  \hline
              \end{tabular}}
        \end{tiny}
    \end{center}
\end{table}

\section{Direct radiation pressure detection}
\label{sec:srp}

Solar radiation pressure is a more complicated perturbation to
detect. So far, solar radiation pressure has only been detected for
very small objects ($H > 27$) who experienced Earth encounters. Thus
we started selecting the smallest asteroids of the initial sample
(more precisely, those with $H>24$) since the effect becomes larger
for smaller size objects, and we tried to detect solar radiation
pressure (SRP) along with the Yarkovsky parameter.

The acceleration caused by solar radiation pressure is radial and can
be modelled with a single parameter $A_1$,
\[
  \mathbf{a}_r = A_1g(r)\hat{\mathbf{r}}.
\]
In this equation $A_1$ is a free parameter, and $g(r)=1/r^2$, where
$r$ is the heliocentric distance in astronomical
  units. Physically, the value of $A_1$ depends mostly on the the
area-to-mass ratio $\mathcal{A}/M$. The relation between them is the
following
\[
    A_1 = \frac{\Phi_\Sun}{c}\cdot C_R \cdot\mathcal{A}/M,
\]
where $c$ is the speed of light, $\Phi_\Sun$ is the solar radiation
energy flux at $1$~au, whose value is $\Phi_\Sun\simeq
1.361$~kW/m$^2$, and $C_R$ is a coefficient (of the order of $1$)
depending upon shape and optical properties of the surface.

\begin{table}[h]
    \setlength{\tabcolsep}{2.5pt}
    \sisetup{table-figures-uncertainty=1}
    \begin{center}
        \caption{List of detections including both the Yarkovsky
          effect and solar radiation pressure, that is the set of
          asteroids for which the parameter $A_1$ was reliably
          determined with a signal-to-noise ratio $\SNR_{A_1}>3$.}
        \label{tab:SRP}
        \vspace{0.3cm}
        \begin{tiny}
          \resizebox{\textwidth}{!}{\begin{tabular}{
                                      p{1.8cm}
                                      S[table-format=5.0,table-column-width=1.2cm]@{\;$\pm$\;}
                                      S[table-format=5.0,table-column-width=1.2cm]
                                      S[table-format=2.1,table-column-width=0.8cm]
                                      S[table-format=1.1,table-column-width=0.8cm]
                                      S[table-format=5.0,table-column-width=1.2cm]@{\;$\pm$\;}
                                      S[table-format=5.0,table-column-width=1.2cm]
                                      S[table-format=1.1,table-column-width=0.8cm]
                                      S[table-format=1.1,table-column-width=1.4cm]
                                      S[table-format=2.0]
                                      }
                  \hline
                  {\bf Asteroid}  & \multicolumn{2}{c}{$A_2$} & {$\text{SNR}_{A_2}$} & {$\mathcal{S}$} & \multicolumn{2}{c}{$A_1$} & {$\text{SNR}_{A_1}$} & {$\mathcal{A}/M$} & {$D$}\\
                              & \multicolumn{2}{c}{($10^{-15}$ au/d$^2$)} &           &              & \multicolumn{2}{c}{($10^{-15}$ au/d$^2$)} &        &  {(m$^2$/ton)}            & {(m)} \\
                  \hline

                  2009 BD & -1152 & 82 & 14.0 & 0.2 & 57663 & 8674 & 6.7 & 0.3 & 4\\
2012 LA & -4907 & 12832 & 0.4 & 2.2 & 81216 & 16312 & 5.0 & 0.4 & 10*\\
2011 MD & -2006 & 3049 & 0.7 & 0.5 & 75074 & 24396 & 3.1 & 0.3 & 6\\
2015 TC$_{25}$ & -4433 & 2754 & 1.6 & 1.4 & 160079 & 20065 & 8.0 & 0.7 & 3*\\
\hline
2006 RH$_{120}$ & -50469 & 3787 & 13.3 & 9.0 & 124099 & 4747 & 26.1 & 0.6 & 4*\\

                  \hline
              \end{tabular}}
        \end{tiny}
    \end{center}
\end{table}

The starting sample of asteroids for which we attempted an
8-dimensional fit contained \srptot{} objects. We found \srp{}
accepted detections, \emph{i.e.} $\SNR_{A_1}\geq 3$, which are listed
in Table~\ref{tab:SRP}. Notice that for three asteroids of this
category, namely 2011~MD, 2012~LA, and 2015~TC$_{25}$, the Yarkovsky
detection is not significant and thus the $\mathcal{S}$ value, though
above the threshold in one case, does not provide any
information. Concerning the area-to-mass ratio we would compare the
value of $\mathcal{A}/M$ with an expected value, as we do with the
secular semimajor axis drift $da/dt$, but this is not possible for now
since the diameter is very uncertain and the other physical properties
are currently unknown. Other fitted values of the area-to-mass ratio
has already been determined for 2009~BD \citep{micheli:2009BD},
2012~LA \citep{micheli:2012la} and 2011~MD \citep{micheli:2011md},
though without including the Yarkovsky effect in the dynamical
model. Asteroid 2006~RH$_{120}$ is listed separately from the others,
since we consider it spurious, as we explain in what follows.

\paragraph{2006~RH$_{120}$.} This strange detection has already been
discussed in \cite{chesley:yarko}. Our results are very compatible
with those of that paper, and we agree with the motivations provided
to reject this detection. The most likely explanation for the high
transverse acceleration can be the presence of some non-conservative
force, \emph{e.g.} mass-shedding, outgassing or micrometeorite flux,
that can become as relevant as the Yarkovsky effect for objects of
this size. The area-to-mass ratio, which results in a significant
detection, is not compatible with the hypothesis that 2006~RH$_{120}$
is an artificial object.

\section{Comparison with JPL results}
\label{sec:compare_JPL}

As we already mentioned, the JPL database is regularly updated with
the asteroids for which the orbital fit shows evidence of the
Yarkovsky effect. The same is done for solar radiation pressure when
appropriate. The results produced by two independent software are
expected to be different, but compatible. In order to compare them we
compute the relative errors
\[
    \varepsilon_r(A_2) \coloneq \frac{|A_2-A_2^{\JPL}|}{\sigma_{A_2}} \quad\text{and}\quad
    \varepsilon_r^{\JPL}(A_2) \coloneq \frac{|A_2-A_2^{\JPL}|}{\sigma_{A_2, \JPL}},
\]
where the superscript ``\textsc{jpl}'' refers to the JPL solution. To
quantify the difference between the results presented in this paper
and the JPL ones, we use the quantity
\[
    \chi_{A_2} \coloneq
    \frac{|A_2-A_2^{\JPL}|}{\sqrt{\sigma_{A_2}^2+\sigma_{A_2,
    \JPL}^2}}
\]
from \cite[Sec.~7.2]{milani:orbdet}. We consider compatible two
solutions for which $\chi_{A_2}\leq 1$.

\begin{table}[h]
    \setlength{\tabcolsep}{2.5pt}
    \sisetup{table-figures-uncertainty=1}
    \begin{center}
        \caption{Comparison between the accepted results of this paper
          ($\SNR_{A_2} \geq 5$) and the JPL ones. The columns contain
          the asteroid name, the signal-to-noise ratio of our solution
          and of the JPL one, the ratio $\sigma_{A_2,
            \JPL}/\sigma_{A_2}$ of the $A_2$ uncertainties as
          estimated by the two systems, the relative errors computed
          with our $A_2$ uncertainty and with the JPL $A_2$
          uncertainty respectively, and the $\chi_{A_2}$ value.}
        \label{tab:compare1}
        \vspace{0.3cm}
        \begin{tiny}
            \begin{tabular}{
                    p{3.5cm}
                    S[table-format=2.1,table-column-width=1.2cm]
                    S[table-format=2.1,table-column-width=1.2cm]
                    S[table-format=1.2,table-column-width=1.8cm]
                    S[table-format=1.2,table-column-width=1.2cm]
                    S[table-format=1.2,table-column-width=1.2cm]
                    S[table-format=1.3,table-column-width=1.2cm]
                    S[table-format=1.2,table-column-width=0.2cm]
                  }
                  \hline
                  {\bf Asteroid} & {$\text{SNR}_{A_2}$} & {$\text{SNR}_{A_2}^{\JPL}$} &
                  {$\sigma_{A_2, \JPL}/\sigma_{A_2}$} & {$\varepsilon_r(A_2)$} & {$\varepsilon_r^{\JPL}(A_2)$} & {$\chi_{A_2}$} &\\

                  \hline

                  (101955) Bennu  & 192.50 & 182.10 & 1.06 & 0.23 & 0.22 & 0.173 & ~\\
(480883) 2001 YE$_{4}$  & 114.54 & 72.38 & 1.58 & 0.29 & 0.18 & 0.149 & ~\\
(2340) Hathor  & 25.37 & 24.29 & 1.06 & 0.29 & 0.28 & 0.204 & ~\\
(483656) 2005 ES$_{70}$  & 25.08 & 18.39 & 1.39 & 0.40 & 0.29 & 0.236 & ~\\
(152563) 1992 BF  & 21.24 & 27.49 & 0.81 & 1.05 & 1.30 & 0.816 & ~\\
2012 BB$_{124}$  & 17.57 & 9.00 & 1.86 & 0.83 & 0.45 & 0.392 & ~\\
(85990) 1999 JV$_{6}$  & 13.98 & 12.58 & 1.22 & 1.37 & 1.12 & 0.869 & ~\\
(437844) 1999 MN  & 10.46 & 8.42 & 1.17 & 0.59 & 0.50 & 0.381 & ~\\
(480808) 1994 XL$_{1}$  & 10.37 & 11.81 & 0.93 & 0.61 & 0.66 & 0.446 & ~\\
2007 TF$_{68}$  & 10.28 & 6.02 & 1.55 & 0.95 & 0.62 & 0.517 & ~\\
(1566) Icarus  & 9.62 & 3.79 & 2.10 & 1.64 & 0.78 & 0.705 & ~\\
(138175) 2000 EE$_{104}$  & 8.96 & 6.86 & 1.20 & 0.72 & 0.60 & 0.460 & ~\\
(1862) Apollo  & 8.81 & 7.23 & 1.12 & 0.71 & 0.63 & 0.476 & ~\\
(2062) Aten  & 8.61 & 7.34 & 1.07 & 0.79 & 0.74 & 0.541 & ~\\
(468468) 2004 KH$_{17}$  & 8.15 & 6.55 & 1.28 & 0.27 & 0.21 & 0.167 & ~\\
(162004) 1991 VE  & 8.05 & 6.10 & 1.14 & 1.07 & 0.93 & 0.704 & ~\\
2006 TU$_{7}$  & 7.73 & 5.58 & 1.40 & 0.08 & 0.06 & 0.045 & ~\\
2011 PU$_{1}$  & 7.56 & 6.00 & 1.01 & 1.52 & 1.51 & 1.074 & $\star$\\
(6489) Golevka  & 7.21 & 7.91 & 0.87 & 0.32 & 0.36 & 0.239 & ~\\
2011 EP$_{51}$  & 7.01 & 6.46 & 0.98 & 0.71 & 0.72 & 0.505 & ~\\
(33342) 1998 WT$_{24}$  & 6.88 & 5.27 & 1.22 & 0.43 & 0.35 & 0.273 & ~\\
(3361) Orpheus  & 6.77 & 7.10 & 1.09 & 0.99 & 0.91 & 0.668 & ~\\
(364136) 2006 CJ  & 6.45 & 8.26 & 0.74 & 0.32 & 0.44 & 0.261 & ~\\
(499998) 2011 PT  & 6.32 & 7.40 & 0.81 & 0.30 & 0.37 & 0.236 & ~\\
(138404) 2000 HA$_{24}$  & 6.30 & 2.05 & 2.82 & 0.53 & 0.19 & 0.177 & ~\\
2006 CT  & 6.22 & 5.82 & 0.99 & 0.45 & 0.45 & 0.319 & ~\\
(3908) Nyx  & 6.06 & 4.62 & 1.29 & 0.08 & 0.06 & 0.047 & ~\\
(363599) 2004 FG$_{11}$  & 5.89 & 3.81 & 1.56 & 0.05 & 0.03 & 0.027 & ~\\
1999 UQ  & 5.88 & 3.39 & 1.87 & 0.43 & 0.23 & 0.205 & ~\\
2003 YL$_{118}$  & 5.87 & 4.71 & 1.18 & 0.32 & 0.28 & 0.210 & ~\\
(154590) 2003 MA$_{3}$  & 5.87 & 4.75 & 1.19 & 0.25 & 0.21 & 0.159 & ~\\
2005 EY$_{169}$  & 5.78 & 4.29 & 1.22 & 0.57 & 0.47 & 0.360 & ~\\
(10302) 1989 ML  & 5.73 & 4.58 & 1.02 & 1.06 & 1.03 & 0.738 & ~\\
2000 PN$_{8}$  & 5.56 & 5.43 & 1.07 & 0.25 & 0.23 & 0.172 & ~\\
(506590) 2005 XB$_{1}$  & 5.28 & 5.59 & 1.09 & 0.78 & 0.72 & 0.531 & ~\\
(216523) 2001 HY$_{7}$  & 5.21 & 4.52 & 1.09 & 0.29 & 0.27 & 0.198 & ~\\
(350462) 1998 KG$_{3}$  & 5.20 & 5.75 & 0.89 & 0.11 & 0.12 & 0.079 & ~\\
(363505) 2003 UC$_{20}$  & 5.19 & 2.57 & 1.22 & 2.06 & 1.68 & 1.302 & $\star$\\
(99907) 1989 VA  & 5.18 & 3.63 & 1.30 & 0.45 & 0.34 & 0.273 & ~\\
(66400) 1999 LT$_{7}$  & 5.17 & 4.37 & 1.20 & 0.06 & 0.05 & 0.037 & ~\\
(377097) 2002 WQ$_{4}$  & 5.13 & 3.93 & 1.32 & 0.05 & 0.04 & 0.029 & ~\\
2000 CK$_{59}$  & 5.10 & 5.76 & 0.87 & 0.10 & 0.11 & 0.074 & ~\\

                  \hline
              \end{tabular}
          \end{tiny}
    \end{center}
\end{table}

\begin{table}[h]
    \setlength{\tabcolsep}{2.5pt}
    \sisetup{table-figures-uncertainty=1}
    \begin{center}
        \caption{Comparison between the accepted results of this paper
          ($\SNR_{A_2} < 5$) and the JPL ones. The columns are the
          same of Table~\ref{tab:compare1}.}
        \label{tab:compare2}
        \vspace{0.3cm}
        \begin{tiny}
            \begin{tabular}{
                    p{3cm}
                    S[table-format=2.1,table-column-width=1.2cm]
                    S[table-format=2.1,table-column-width=1.2cm]
                    S[table-format=1.2,table-column-width=1.8cm]
                    S[table-format=1.2,table-column-width=1.2cm]
                    S[table-format=1.2,table-column-width=1.2cm]
                    S[table-format=1.3,table-column-width=1.2cm]
                    S[table-format=1.2,table-column-width=0.2cm]
                  }
                  \hline
                  {\bf Asteroid} & {$\text{SNR}_{A_2}$} & {$\text{SNR}_{A_2}^{\JPL}$} &
                  {$\sigma_{A_2, \JPL}/\sigma_{A_2}$} & {$\varepsilon_r(A_2)$} & {$\varepsilon_r^{\JPL}(A_2)$} & {$\chi_{A_2}$} &\\

                  \hline

                  (29075) 1950 DA  & 4.82 & 4.17 & 1.03 & 0.51 & 0.49 & 0.351 & ~\\
(162117) 1998 SD$_{15}$  & 4.74 & 4.35 & 1.23 & 0.62 & 0.51 & 0.394 & ~\\
2001 BB$_{16}$  & 4.68 & 4.86 & 1.14 & 0.86 & 0.75 & 0.567 & ~\\
(138852) 2000 WN$_{10}$  & 4.62 & 4.29 & 1.04 & 0.17 & 0.16 & 0.116 & ~\\
(455176) 1999 VF$_{22}$  & 4.55 & 3.99 & 1.27 & 0.51 & 0.40 & 0.313 & ~\\
(399308) 1993 GD  & 4.51 & 4.45 & 0.98 & 0.17 & 0.17 & 0.120 & ~\\
(1685) Toro  & 4.48 & 4.33 & 0.85 & 0.81 & 0.95 & 0.618 & ~\\
(7336) Saunders  & 4.46 & 2.76 & 1.22 & 1.08 & 0.89 & 0.686 & ~\\
(4034) Vishnu  & 4.28 & 4.31 & 1.15 & 0.66 & 0.57 & 0.433 & ~\\
(85774) 1998 UT$_{18}$  & 4.28 & 3.44 & 1.11 & 0.48 & 0.44 & 0.325 & ~\\
(310442) 2000 CH$_{59}$  & 4.26 & 2.60 & 1.38 & 0.68 & 0.50 & 0.402 & ~\\
(2100) Ra-Shalom  & 4.23 & 3.10 & 1.14 & 0.69 & 0.61 & 0.456 & ~\\
(326354) 2000 SJ$_{344}$  & 4.20 & 6.92 & 0.65 & 0.28 & 0.44 & 0.237 & ~\\
(481442) 2006 WO$_{3}$  & 4.15 & 3.97 & 0.94 & 0.41 & 0.44 & 0.300 & ~\\
(441987) 2010 NY$_{65}$  & 4.08 & 3.88 & 1.03 & 0.07 & 0.07 & 0.047 & ~\\
(306383) 1993 VD  & 4.08 & 1.61 & 1.35 & 1.91 & 1.42 & 1.137 & $\star$\\
2008 CE$_{119}$  & 3.98 & 3.43 & 1.57 & 1.41 & 0.90 & 0.756 & ~\\
(85953) 1999 FK$_{21}$  & 3.96 & 4.94 & 0.89 & 0.44 & 0.49 & 0.327 & ~\\
(348306) 2005 AY$_{28}$  & 3.94 & 4.23 & 0.75 & 0.76 & 1.01 & 0.606 & ~\\
(65679) 1989 UQ  & 3.86 & 3.74 & 1.07 & 0.14 & 0.13 & 0.097 & ~\\
(232691) 2004 AR$_{1}$  & 3.83 & 2.28 & 1.08 & 1.38 & 1.29 & 0.942 & ~\\
1995 CR  & 3.83 & 2.23 & 1.42 & 0.65 & 0.46 & 0.374 & ~\\
(265482) 2005 EE  & 3.82 & 1.58 & 1.48 & 1.46 & 0.99 & 0.818 & ~\\
(136818) Selqet  & 3.81 & 2.08 & 1.37 & 0.96 & 0.70 & 0.566 & ~\\
(425755) 2011 CP$_{4}$  & 3.76 & 3.37 & 1.24 & 0.40 & 0.33 & 0.254 & ~\\
(192559) 1998 VO  & 3.75 & 3.77 & 0.92 & 0.26 & 0.29 & 0.194 & ~\\
(163023) 2001 XU$_{1}$  & 3.72 & 2.94 & 1.05 & 0.63 & 0.60 & 0.432 & ~\\
(397326) 2006 TC$_{1}$  & 3.65 & 3.45 & 0.98 & 0.26 & 0.27 & 0.188 & ~\\
(208023) 1999 AQ$_{10}$  & 3.64 & 2.27 & 1.14 & 1.05 & 0.93 & 0.696 & ~\\
(5604) 1992 FE  & 3.64 & 3.46 & 1.24 & 0.64 & 0.52 & 0.402 & ~\\
(437841) 1998 HD$_{14}$  & 3.58 & 3.26 & 0.87 & 0.73 & 0.84 & 0.551 & ~\\
(413260) 2003 TL$_{4}$  & 3.53 & 2.85 & 1.02 & 0.61 & 0.60 & 0.429 & ~\\
(4581) Asclepius  & 3.47 & 2.57 & 1.20 & 0.39 & 0.32 & 0.247 & ~\\
(467351) 2003 KO$_{2}$  & 3.44 & 2.93 & 1.34 & 0.49 & 0.37 & 0.294 & ~\\
(136582) 1992 BA  & 3.36 & 3.43 & 1.18 & 0.70 & 0.59 & 0.449 & ~\\
(256004) 2006 UP  & 3.28 & 3.65 & 0.95 & 0.18 & 0.19 & 0.133 & ~\\
(7341) 1991 VK  & 3.28 & 3.72 & 1.07 & 0.68 & 0.64 & 0.465 & ~\\
(450300) 2004 QD$_{14}$  & 3.26 & 1.99 & 2.03 & 0.78 & 0.38 & 0.342 & ~\\
(477719) 2010 SG$_{15}$  & 3.19 & 2.78 & 1.01 & 0.39 & 0.38 & 0.272 & ~\\
(37655) Illapa  & 3.15 & 2.67 & 1.14 & 0.10 & 0.09 & 0.065 & ~\\
(267759) 2003 MC$_{7}$  & 3.12 & 3.57 & 0.87 & 0.01 & 0.02 & 0.010 & ~\\
(310842) 2003 AK$_{18}$  & 3.06 & 2.60 & 1.31 & 0.35 & 0.27 & 0.214 & ~\\
(162783) 2000 YJ$_{11}$  & 3.02 & 3.38 & 0.94 & 0.16 & 0.17 & 0.114 & ~\\
(152671) 1998 HL$_{3}$  & 3.02 & 3.16 & 0.98 & 0.06 & 0.06 & 0.042 & ~\\
(85770) 1998 UP$_{1}$  & 2.94 & 3.01 & 1.42 & 1.34 & 0.94 & 0.771 & ~\\
(474163) 1999 SO$_{5}$  & 2.88 & 3.51 & 1.02 & 0.72 & 0.70 & 0.501 & ~\\
(283457) 2001 MQ$_{3}$  & 2.80 & 3.91 & 0.87 & 0.61 & 0.70 & 0.462 & ~\\
(376879) 2001 WW$_{1}$  & 2.57 & 3.07 & 0.76 & 0.24 & 0.31 & 0.188 & ~\\
(99942) Apophis  & 1.80 & 2.54 & 0.74 & 0.09 & 0.12 & 0.069 & ~\\
(410777) 2009 FD  & 0.45 & 0.04 & 1.22 & 0.51 & 0.42 & 0.321 & ~\\

                  \hline
              \end{tabular}
          \end{tiny}
    \end{center}
\end{table}

Starting from the list of our accepted detections, we compared the
results every time an asteroid is included in the JPL database of
Yarkovsky effect detections. The results of the comparison are
contained in Table~\ref{tab:compare1} and~\ref{tab:compare2}. Just for
five asteroids in this list both the relative errors are greater than
1, even though never above $2.5$. Using the metric given by
$\chi_{A_2}$, we identify just $3$ asteroids (marked with a star in
Table~\ref{tab:compare1} and~\ref{tab:compare2}) whose detections are
not fully compatible with respect to our criteria. Anyway, a
$\chi_{A_2}$ moderately above $1$ for $3$ cases out of $92$ being
compared shows a strong agreement between our results and the JPL's
ones.

Note that this result is not a null test, that is the expected value
of the difference in the estimated values of $A_2$ is not zero. This
because the two computations have used two different astrometric error
models, \cite{farnocchia:fcct} at NEODyS and \cite{veres:2017} at
JPL. The comparative results described in the last three columns
indicate an exceptionally good agreement. This agreement may be
interpreted as a validation of the procedures used both at NEODyS and
at JPL, both to compute the Yarkovsky effect constants and to select
the cases in which the results are reliable.

The comparison was also performed for the shorter list of objects for
which we have both $A_2$ and $A_1$, that is both Yarkovsky effect and
direct radiation pressure were included in the dynamical
model. Table~\ref{tab:compare_JPL_A1A2} contains the signal-to-noise
ratios for both parameters in both solutions, and all the metrics for
the discrepancies. Apart from the results for 2006~RH$_{120}$, which
are rated as spurious, the accepted results are fully consistent.

\begin{table}[h]
    \setlength{\tabcolsep}{2.5pt}
    \sisetup{table-figures-uncertainty=1}
    \begin{center}
        \caption{Results of the comparison between the estimated
          values of $A_2$ and $A_1$, as contained in this paper and in
          the JPL database. In particular, the columns contain the
          asteroid name, the signal-to-noise ratio of our $A_2$
          solution and of the JPL one, the signal-to-noise ratio of
          our $A_1$ solution and of the JPL one, the relative error in
          the $A_2$ parameter computed with our $A_2$ uncertainty and
          with the JPL $A_2$ uncertainty respectively, the relative
          error in the $A_1$ parameter computed with our $A_1$
          uncertainty and with the JPL $A_1$ uncertainty respectively,
          the $\chi$-value for $A_2$ and for $A_1$.}
        \label{tab:compare_JPL_A1A2}
        \vspace{0.3cm}
        \begin{tiny}
            \begin{tabular}{
                  p{1.4cm}
                  S[table-format=2.1,table-column-width=0.9cm]
                  S[table-format=2.1,table-column-width=0.9cm]
                  S[table-format=2.1,table-column-width=1.0cm]
                  S[table-format=2.1,table-column-width=1.0cm]
                  S[table-format=1.2,table-column-width=1.0cm]
                  S[table-format=1.2,table-column-width=1.0cm]
                  S[table-format=1.2,table-column-width=1.0cm]
                  S[table-format=1.2,table-column-width=1.0cm]
                  S[table-format=1.3,table-column-width=0.8cm]
                  S[table-format=1.3,table-column-width=0.8cm]
                }
                \hline
                {\bf Asteroid} & {$\text{SNR}_{A_2}$} & {$\text{SNR}_{A_2}^{\JPL}$} & {$\text{SNR}_{A_1}$} & {$\text{SNR}_{A_1}^{\JPL}$} & {$\varepsilon_r(A_2)$} & {$\varepsilon_r^{\JPL}(A_2)$} & {$\varepsilon_r(A_1)$} & {$\varepsilon_r^{\JPL}(A_1)$} & {$\chi_{A_2}$} & {$\chi_{A_1}$}\\

                \hline

                2009 BD  & 14.0 & 13.9 & 6.7 & 6.3 & 0.12 & 0.12 & 0.44 & 0.45 & 0.084 & 0.315\\
2012 LA  & 0.4 & 0.3 & 5.0 & 6.9 & 0.22 & 0.37 & 0.04 & 0.05 & 0.189 & 0.029\\
2011 MD  & 0.7 & 0.3 & 3.1 & 3.1 & 0.32 & 0.25 & 0.11 & 0.10 & 0.198 & 0.074\\
2015 TC$_{25}$  & 1.6 & 1.6 & 8.0 & 7.6 & 0.06 & 0.05 & 0.11 & 0.11 & 0.039 & 0.076\\
\hline
2006 RH$_{120}$  & 13.3 & 11.1 & 26.1 & 23.4 & 0.16 & 0.13 & 1.62 & 1.36 & 0.100 & 1.043\\

                \hline
            \end{tabular}
        \end{tiny}
    \end{center}
\end{table}

\section{Impact monitoring with non-gravitational parameters}
\label{sec:im_nongrav}

A force model including non-gravitational forces is sometimes needed
to make reliable impact predictions, especially if we want to extend
the hazard analysis time span to longer intervals with respect to one
century (the default time span adopted by the current impact
monitoring systems). More precisely, the non-gravitational model plays
a fundamental role also in the Line Of Variations (LOV) computation
and propagation
\citep{milani:multsol,milani:clomon2,milani:1999risk}. If an asteroid
with a very well constrained orbit experiences a very deep close
approach, the post-encounter situation is equal to the one of a poorly
determined orbit, with the difference that the large uncertainty of
the asteroid state is due to the divergence of nearby orbits caused by
the encounter, and not to the poor constraints of the initial
conditions. In this case, the initial confidence region is very small,
thus the use of the linear approximation of the LOV is allowed. In
case such an encounter occurs the linear LOV direction is derived by
analyzing that encounter and mapping back the corresponding LOV trace
on the target plane (TP) \citep{valsecchi:resret} to the space of
initial conditions. This method has been used in \cite{spoto:410777}
to properly assess the impact risk of (410777)~2009~FD, exploiting its
2185 scattering encounter with the Earth. The same formalism can be
used even when we are not in the presence of a scattering encounter,
but the close encounter is so deep that the LOV will turn out to be
quite stretched at the next encounter, as in the cases analyzed below.

So far, just four asteroids required such special treatment for a
proper impact risk assessment, namely (101955)~Bennu, (99942)~Apophis,
(29075) 1950~DA, and (410777)~2009~FD, but this list is expected to
grow as a consequence of the work presented in this paper. Below we
show two examples of asteroids for which we found virtual impactors
using a non-gravitational model and that have no possible impacts with
a purely gravitational model. We are aware that such a work could be
done on many asteroids with accepted Yarkovsky detections, but this is
beyond the scope of this paper.

\paragraph{2001~BB$_{16}$.} Currently, this asteroid has a low MOID
value, $\simeq 0.0043$~au, but no chance of impacting the Earth in the
next century. 2001~BB$_{16}$ has a deep close approach with the Earth
in 2082, which causes an increase of the stretching of two orders of
magnitude with respect to the next 2086 encounter, whereas the
stretching value remains essentially constant until the 2082 close
approach. We used this close approach to derive the LOV direction and
we performed the impact monitoring through 2200 employing a
non-gravitational model including the Yarkovsky effect. The results
are shown in Table~\ref{tab:2001BB16}. In particular we found two VIs
at the very end of the 22nd century, which we would not find with a
gravity-only model.

\begin{table}[h]
    \setlength{\tabcolsep}{2.5pt}
    \sisetup{table-figures-uncertainty=1}
    \begin{center}
        \caption{Impact monitoring of asteroid 2001~BB$_{16}$ with a
          non-gravitational model that includes the Yarkovsky
          effect. Table columns: calendar date (year, month, and day)
          for the potential impact for asteroid 2011~MD, approximate
          $\sigma$ value of the virtual impactor location along the
          LOV, minimum distance (the lateral distance from the LOV to
          the center of the Earth on the TP confidence region),
          stretching (how much the confidence region at the epoch has
          been stretched by the time of impact), probability of Earth
          impact ($IP$), and Palermo Scale ($PS$). The width of the TP
          confidence region is always few km, thus not reported.}
        \label{tab:2001BB16}
        \vspace{0.3cm}
        \begin{tiny}
            \begin{tabular}{
                  c
                  S[table-format=2.3,table-column-width=1.2cm]
                  c
                  c
                  c
                  c
                }
                \hline
                {\bf Date} & {$\sigma$} & {\bf Distance} & {\bf Stretching} & {$IP$} & {$PS$} \\
                &           & $(R_\Earth)$ &   $(R_\Earth)$  &        &        \\
                \hline

                2195/01/15.525  & -3.404 & 3.17 & $ 5.20 \cdot 10^{3} $ & $ 2.91 \cdot 10^{-7} $ & ${-6.35}$\\
2199/01/15.844  & -3.164 & 2.10 & $ 8.36 \cdot 10^{4} $ & $ 5.02 \cdot 10^{-8} $ & ${-7.12}$\\

                \hline
            \end{tabular}
        \end{tiny}
    \end{center}
\end{table}

\paragraph{2011~MD.} This is a very small asteroid, about $6$~m in
diameter, as determined in \cite{mommert:2011MD}. In this case as
well, the MOID value is very low, $\simeq 0.00036$~au and it has no
virtual impactor in the next century. In 2049, this asteroid will
experience two very close approaches with the Earth, causing an
increase of two orders of magnitude in the stretching between these
encounters and the following one in 2067. We used the first 2049 close
approach (the deepest of the two) to compute the LOV direction in the
space of initial conditions. We thus performed the impact monitoring
using a dynamical model including both the Yarkovsky effect and solar
radiation pressure. The results are shown in
Table~\ref{tab:2011MD_1}. When we only include solar radiation
pressure, the orbit uncertainty shrinks and thus the number of VIs is
much lower than before (see Table~\ref{tab:2011MD_2}). Both
Table~\ref{tab:2011MD_1} and Table~\ref{tab:2011MD_2} list the virtual
impactors with $IP\ge 10^{-7}$, since this threshold is the
completeness limit used for the LOV sampling
\citep{delvigna:compl_IM}.

\noindent It is worth noting that this asteroid is so small that it
would not reach the Earth in case of a real impact, because it would
be burnt in the atmosphere. This case is studied to show that,
  in some cases, a non-gravitational model is needed to make reliable
  impact predictions and also that different models of
  non-gravitational perturbations can give very different results.

\begin{table}[h]
    \setlength{\tabcolsep}{2.5pt}
    \sisetup{table-figures-uncertainty=1}
    \begin{center}
        \caption{Impact monitoring of asteroid 2011~MD with a
          non-gravitational model that includes both the Yarkovsky
          effect and solar radiation pressure. Columns as in
          Table~\ref{tab:2001BB16}.}
        \label{tab:2011MD_1}
        \vspace{0.3cm}
        \begin{tiny}
            \begin{tabular}{
                  c
                  S[table-format=2.3,table-column-width=1.2cm]
                  c
                  c
                  c
                  c
                }
                \hline
                {\bf Date} & {$\sigma$} & {\bf Distance} & {\bf Stretching} & {$IP$} & {$PS$} \\
                &           & $(R_\Earth)$ &  $(R_\Earth)$   &        &        \\
                \hline

                2083/06/13.856  & 2.720 & 6.44 & $ 2.86 \cdot 10^{4} $ & $ 3.54 \cdot 10^{-7} $ & ${-8.29}$\\
2098/06/07.618  & 0.428 & 6.38 & $ 9.00 \cdot 10^{5} $ & $ 4.33 \cdot 10^{-7} $ & ${-8.29}$\\
2099/06/08.786  & 0.370 & 5.68 & $ 3.70 \cdot 10^{5} $ & $ 1.32 \cdot 10^{-6} $ & ${-7.81}$\\
2102/06/13.699  & -0.250 & 6.40 & $ 5.23 \cdot 10^{4} $ & $ 7.61 \cdot 10^{-6} $ & ${-7.06}$\\
2110/05/28.604  & -0.993 & 5.52 & $ 1.48 \cdot 10^{6} $ & $ 2.09 \cdot 10^{-7} $ & ${-8.66}$\\
2113/06/09.765  & 0.330 & 4.09 & $ 9.48 \cdot 10^{5} $ & $ 6.90 \cdot 10^{-7} $ & ${-8.15}$\\
2116/06/08.850  & 0.258 & 6.37 & $ 2.61 \cdot 10^{6} $ & $ 1.57 \cdot 10^{-7} $ & ${-8.81}$\\
2116/06/08.870  & 0.258 & 6.24 & $ 3.78 \cdot 10^{6} $ & $ 1.14 \cdot 10^{-7} $ & ${-8.95}$\\
2118/06/07.317  & 0.472 & 6.12 & $ 1.52 \cdot 10^{6} $ & $ 2.65 \cdot 10^{-7} $ & ${-8.59}$\\
2118/06/10.605  & 1.496 & 4.80 & $ 3.54 \cdot 10^{5} $ & $ 5.66 \cdot 10^{-7} $ & ${-8.26}$\\
2119/06/14.086  & -0.670 & 6.55 & $ 3.30 \cdot 10^{4} $ & $ 9.22 \cdot 10^{-6} $ & ${-7.05}$\\
2119/06/14.181  & -0.670 & 1.70 & $ 9.79 \cdot 10^{5} $ & $ 6.57 \cdot 10^{-7} $ & ${-8.20}$\\
2120/06/04.434  & 0.520 & 5.51 & $ 3.44 \cdot 10^{6} $ & $ 1.69 \cdot 10^{-7} $ & ${-8.79}$\\
2122/06/11.859  & 1.223 & 5.48 & $ 1.25 \cdot 10^{6} $ & $ 2.07 \cdot 10^{-7} $ & ${-8.72}$\\
2123/06/08.803  & 0.416 & 4.78 & $ 1.28 \cdot 10^{6} $ & $ 4.34 \cdot 10^{-7} $ & ${-8.40}$\\
2132/06/09.046  & 0.184 & 6.30 & $ 3.78 \cdot 10^{6} $ & $ 1.13 \cdot 10^{-7} $ & ${-9.02}$\\
2137/06/11.124  & 1.218 & 6.18 & $ 2.09 \cdot 10^{6} $ & $ 1.01 \cdot 10^{-7} $ & ${-9.08}$\\
2139/06/12.675  & 1.240 & 5.83 & $ 8.80 \cdot 10^{5} $ & $ 2.58 \cdot 10^{-7} $ & ${-8.68}$\\
2140/06/12.052  & 1.301 & 4.52 & $ 7.04 \cdot 10^{5} $ & $ 3.98 \cdot 10^{-7} $ & ${-8.50}$\\
2151/06/12.350  & -0.796 & 6.93 & $ 1.66 \cdot 10^{6} $ & $ 1.20 \cdot 10^{-7} $ & ${-9.05}$\\
2151/06/12.604  & -0.732 & 6.44 & $ 2.90 \cdot 10^{6} $ & $ 1.05 \cdot 10^{-7} $ & ${-9.11}$\\
2151/06/12.780  & -0.762 & 2.96 & $ 3.16 \cdot 10^{6} $ & $ 1.76 \cdot 10^{-7} $ & ${-8.89}$\\
2155/06/11.940  & -0.439 & 6.54 & $ 5.27 \cdot 10^{4} $ & $ 5.87 \cdot 10^{-6} $ & ${-7.37}$\\
2155/06/11.984  & -0.436 & 2.86 & $ 2.98 \cdot 10^{5} $ & $ 2.29 \cdot 10^{-6} $ & ${-7.78}$\\
2155/06/12.054  & -0.439 & 2.92 & $ 8.38 \cdot 10^{4} $ & $ 8.04 \cdot 10^{-6} $ & ${-7.24}$\\
2155/06/12.168  & -0.437 & 7.06 & $ 1.96 \cdot 10^{4} $ & $ 1.44 \cdot 10^{-5} $ & ${-6.98}$\\
2158/06/11.842  & -2.091 & 7.19 & $ 2.80 \cdot 10^{5} $ & $ 1.19 \cdot 10^{-7} $ & ${-9.08}$\\
2182/06/09.849  & 0.041 & 4.51 & $ 5.95 \cdot 10^{6} $ & $ 1.09 \cdot 10^{-7} $ & ${-9.18}$\\

                \hline
            \end{tabular}
        \end{tiny}
    \end{center}
\end{table}

\begin{table}[h]
    \setlength{\tabcolsep}{2.5pt}
    \sisetup{table-figures-uncertainty=1}
    \begin{center}
        \caption{Impact monitoring of asteroid 2011~MD with a
          non-gravitational model including solar radiation pressure
          only. Columns as in Table~\ref{tab:2001BB16}.}
        \label{tab:2011MD_2}
        \vspace{0.3cm}
        \begin{tiny}
            \begin{tabular}{
                  c
                  S[table-format=2.3,table-column-width=1.2cm]
                  c
                  c
                  c
                  c
                }
                \hline
                {\bf Date} & {$\sigma$} & {\bf Distance} & {\bf Stretching} & {$IP$} & {$PS$} \\
                &           & $(R_\Earth)$ &  $(R_\Earth)$    &        &        \\
                \hline

                2133/12/05.197  & -0.476 & 0.27 & $ 1.98 \cdot 10^{6} $ & $ 3.58 \cdot 10^{-7} $ & ${-8.52}$\\
2140/11/25.578  & -0.352 & 4.66 & $ 5.53 \cdot 10^{6} $ & $ 1.12 \cdot 10^{-7} $ & ${-9.05}$\\
2147/11/27.042  & 0.212 & 4.79 & $ 1.17 \cdot 10^{5} $ & $ 5.13 \cdot 10^{-6} $ & ${-7.41}$\\
2168/05/22.293  & 0.081 & 4.70 & $ 2.13 \cdot 10^{6} $ & $ 2.87 \cdot 10^{-7} $ & ${-8.72}$\\
2169/11/26.849  & -0.053 & 5.48 & $ 1.92 \cdot 10^{6} $ & $ 2.76 \cdot 10^{-7} $ & ${-8.74}$\\
2186/11/21.935  & -0.633 & 6.87 & $ 1.56 \cdot 10^{6} $ & $ 1.46 \cdot 10^{-7} $ & ${-9.06}$\\

                \hline
            \end{tabular}
        \end{tiny}
    \end{center}
\end{table}

\section{Conclusions and future work}
\label{sec:conclusion}

In this paper we significantly increased the knowledge of
non-gravitational perturbations on near-Earth asteroids, based on
actual measurements, rather than on modelling. The number of
significant and reliable Yarkovsky detections in the NEA catalog is
expected to grow continuously. In fact, the data volume of future
surveys, the increased astrometric accuracy for optical observations,
more accurate star catalog debiasing techniques, and expanded efforts
in radar astrometry provide ever better constraints to measure this
small effect. We identified \yarko{} near-Earth asteroids with
significant and reliable Yarkovsky detection, thus doubling the list
provided in \cite{chesley:yarko}. For few exceedingly small asteroids,
we attempted to directly detect solar radiation pressure together with
the Yarkovsky-related acceleration. For such cases, solar radiation
pressure is needed to obtain a more satisfactory orbital fit.

There are several research centers handling the computation of
asteroid orbits as an industrial production, like recomputing either
all the orbits of more than $500,000$ numbered asteroids every time a
change in the error model occurs, or a large portion of them just to
take into account new observations and new asteroid
discoveries\footnote{The authors of this paper all belong to four
  centers performing this kind of activity: NEODyS, JPL, IMCCE,
  NEOCC.}. There are important scientific goals such as asteroid
families and impact monitoring that can only be achieved by
maintaining and constantly updating such large lists of orbits.

We dedicated a significant effort in clarifying a number of marginal
and/or spurious cases, not only to recover few dubious cases but also
to refine the methodology and therefore be ready for the future
increase of significant detections. Indeed, the problem to be faced in
the near future is not another increase by a factor two, rather an
increase by orders of magnitude. The second Gaia data release (April
2018) will contain about 1.7~billion of sources brighter than
magnitude 21 and $\simeq 14000$ asteroids with astrometry reaching the
sub-milliarcsec accuracy in an optimal range of magnitude $G\simeq
12-17$ \citep{brown:gaia2,spoto:gaia2}. The stellar catalog produced
by Gaia will represent the starting point for a new debiasing and
weighting scheme. Moreover, the combination of Gaia asteroid
observations with the already available ones will produce a sharp
increase in the number of objects for which the Yarkovsky effect will
be detectable. Thus the challenge in papers like this is not to
establish a new record list of Yarkovsky and/or radiation pressure
detections, but rather to develop an automated calculation of orbits
with estimated non-gravitational parameters.

The computations of orbits with non-gravitational effects is still
very far from being an automated process. To avoid spurious
detections, we used the most recent error models for the observations
and a filtering criterion, based on an estimate of the Yarkovsky
effect based upon a physical model of the asteroid. Unfortunately,
both of these tools are still incomplete. The error models suffer from
the continued unavailability of metadata, such as the signal-to-noise
of individual observations, with the result that observations with
different quality are bundled together and the statistical analysis of
the residuals does not yet allow a correct derivation of uncertainty
of the measurement error. The physical models of asteroids, needed to
estimate the expected Yarkovsky effect, are very rough approximations
when the main physical data are not available, as it is the case for
the majority of the asteroids in our tables. Moreover, such small
perturbations can be sensitive to old isolated, and possibly bad
astrometric positions.

In conclusion, we made a step in the right direction by developing and
testing the use of different error models, and by using the difference
in the results as an estimate of the sensitivity of the results on the
error model. We developed and tested the use of a filter for spurious
cases, which is based on an estimate of the expected Yarkovsky effect,
which is roughly the same as the Yarkovsky calibration used to compute
the age of asteroid families
\cite{milani:fam_class,spoto:fam_ages}. Both tools improved our
capability of obtaining a list of reliable Yarkovsky detections, as
well as a much shorter list of radiation pressure detections for
natural bodies.

\section*{Acknowledgements}
We thank the referee Dr. David Vokrouhlicky for his useful comments
that have improved the quality of the paper.

A.~Del Vigna and L.~Faggioli acknowledges support by the company
SpaceDyS. D.~Farnocchia conducted this research at the Jet Propulsion
Laboratory, California Institute of Technology, under a contract with
NASA.

This research was conducted under European Space Agency contract
No. 4000113555/15/DMRP ``P2-NEO-II Improved NEO Data Processing
Capabilities''.

\bibliographystyle{elsarticle-harv}
\bibliography{yarko2_biblio}

\end{document}